\documentclass[aps,pre,superscriptaddress,twocolumn,nofootinbib]{revtex4-1}
\usepackage[utf8]{inputenc}
\usepackage{hyperref}
\usepackage{graphicx}  
\usepackage{dcolumn}   
\usepackage{bm}        
\usepackage{amssymb,amsmath}   
\usepackage{uri}

\usepackage[x11names, rgb, html]{xcolor}
\definecolor{sbase03}{HTML}{002B36}
\definecolor{sbase02}{HTML}{073642}
\definecolor{sbase01}{HTML}{586E75}
\definecolor{sbase00}{HTML}{657B83}
\definecolor{sbase0}{HTML}{839496}
\definecolor{sbase1}{HTML}{93A1A1}
\definecolor{sbase2}{HTML}{EEE8D5}
\definecolor{sbase3}{HTML}{FDF6E3}
\definecolor{syellow}{HTML}{B58900}
\definecolor{sorange}{HTML}{CB4B16}
\definecolor{sred}{HTML}{DC322F}
\definecolor{smagenta}{HTML}{D33682}
\definecolor{sviolet}{HTML}{6C71C4}
\definecolor{sblue}{HTML}{268BD2}
\definecolor{scyan}{HTML}{2AA198}
\definecolor{sgreen}{HTML}{859900}
\hypersetup{
  colorlinks = true,
  citecolor = {sred},
  urlcolor = {sblue}
}

\newcommand{\arcsinh}{\text{arcsinh}}
\newcommand{\csch}{\text{csch}}
\begin{document}
\title{Inferring dissipation from current fluctuations}
\author{Todd R.~Gingrich}
\affiliation{Physics of Living Systems Group, Department of Physics, Massachusetts Institute of Technology, 400 Technology Square, Cambridge, MA 02139}
\email{toddging@mit.edu} 
\author{Grant M.~Rotskoff}
\affiliation{Biophysics Graduate Group, University of California, Berkeley, CA 94720}
\author{Jordan M.~Horowitz}
\affiliation{Physics of Living Systems Group, Department of Physics, Massachusetts Institute of Technology, 400 Technology Square, Cambridge, MA 02139}

\begin{abstract}
Complex physical dynamics can often be modeled as a Markov jump process between mesoscopic configurations.
When jumps between mesoscopic states are mediated by thermodynamic reservoirs, the time-irreversibility of the jump process is a measure of the physical dissipation.
We rederive a recently introduced inequality relating the dissipation rate to current fluctuations in jump processes.
We then adapt these results to diffusion processes via a limiting procedure, reaffirming that diffusions saturate the inequality.
Finally, we study the impact of spatial coarse-graining in a two-dimensional model with driven diffusion.
By observing fluctuations in coarse-grained currents, it is possible to infer a lower bound on the total dissipation rate, including the dissipation associated with hidden dynamics.
The tightness of this bound depends on how well the spatial coarse-graining detects dynamical events that are driven by large thermodynamic forces.
\end{abstract}

\maketitle

\section{Introduction}

Experiments capable of probing molecular-scale dynamics have led to a wealth of data about the operation of nanoscale machines~\cite{Smith1992,Blickle2012,Martinez2016}.
Like their macroscopic counterparts, nanomachines convert free energy from the environment into useful work.
Due to molecular fluctuations, however, nanoscale machines behave stochastically.
Though the stochasticity renders nanomachines less predictable, the fluctuations also offers a unique tool to deduce additional physical properties of the machine.
Indeed, the fluctuation-dissipation theorem (FDT) demonstrates that near-equilibrium fluctuations allow one to deduce the nonequilibrium response to small perturbations, an idea that has become a central theme of statistical mechanics~\cite{Marconi2008, Einstein1905, Onsager1931, Onsager1931II, Callen1951, Kubo1966}.
Analogous exploitation of far-from-equilibrium fluctuations is highly desirable~\cite{Harada2005, Speck2006, Baiesi2009, Prost2009, Seifert2010}, particularly for studying living systems~\cite{Bohec2013,Fodor2015activity, Fodor2016}.

In this work, we demonstrate how it is possible to extract information from nonequilibrium fluctuations under general steady-state conditions.
Our central assumption is that the mesoscale dynamics, in a quantum or a classical system, can be described by a Markovian stochastic process.
For example, the Markovian mesoscopic description applies to biochemical kinetics of enzymatic reactions.
In this context, dynamical fluctuations have been used to constrain proposed models for enzymatic pathways, providing bounds on the number of distinct intermediate states~\cite{Moffitt2014, Barato2015Fano}.
More recently, advances in the theory of Markov processes~\cite{Kesidis1993,Maes2008,Bertini2015Large,Bertini2015Flows} have been used to relate the dynamical fluctuations to the physical dissipation rate of the nonequilibrium dynamics~\cite{Barato2015,Pietzonka2016universal,Gingrich2016,Polettini2016}.
This paper reviews and extends that connection between fluctuations and dissipation, a connection which can be thought of as a generalization of the FDT.
Where the FDT offers an equality\textemdash by measuring fluctuations we may determine the dissipation exactly~\cite{Kubo2012}\textemdash the far-from-equilibrium analog guarantees an inequality: the extent of fluctuations sets a lower bound on the dissipation rate.
Hence experimental observations of macroscopic, coarse-grained fluctuations provide enough information to estimate a bound on the dissipation rate of a process far from equilibrium.

In this paper, we study this generalization of the FDT bound from the perspective of coarse-graining, a viewpoint which we anticipate will complement experimental applications.
After reviewing the formalism of Markov jump processes in Section~\ref{sec:graph}, we consider the consequence of temporal coarse-graining in Section~\ref{sec:heuristic}.
By focusing on dynamical quantities which are time-averaged over long trajectories, we arrive at the so-called Level 2.5 large deviations, which quantify dynamical fluctuations in the jump process~\cite{Kesidis1993,Maes2008,Bertini2015Large,Bertini2015Flows}.
Mirroring our previous work~\cite{Gingrich2016}, Section~\ref{sec:ratefunctionbound} develops our central relationship between nonequilibrium dynamical fluctuations and dissipation.
In its most general form, the relationship pertains to a fluctuating vector quantity in the space of all microscopic currents; the implication for scalar currents is detailed in Section~\ref{sec:scalarcurrent}.
In addition to the temporal coarse-graining, we investigate the role of spatial resolution by studying two limits: a diffusive limit in Section~\ref{sec:diffusion} and the spatial coarse-graining of a particular diffusion process in Section~\ref{sec:macrostate}.
A consequence of the coarse-graining is that the macrostate dynamics may not be Markovian.
Remarkably, even without the Markov property, coarse-grained fluctuations can bound the dissipation rate so long as the macroscopic dynamics emerges from an underlying Markovian description, suggesting that suitable analysis of dynamical fluctuations can complement existing methods for deducing dissipation from time series data~\cite{Roldan2010}.

\section{Markov Jumps on Graphs}
\label{sec:graph}
\subsection{Dissipation}
\begin{figure*}
\centering
\includegraphics[width=0.85\textwidth]{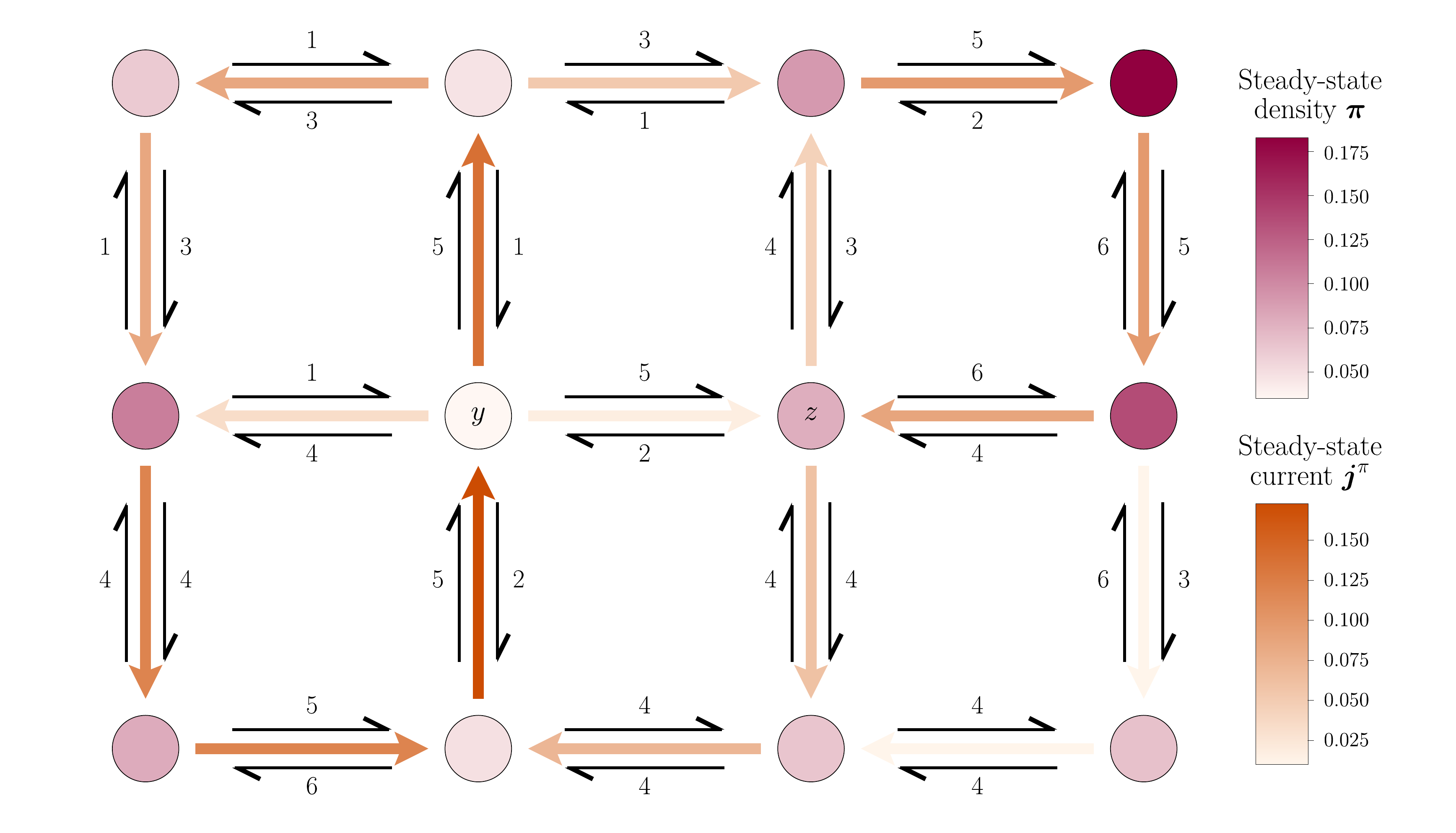}
\caption{Example of a Markov jump process: Mesostates are shaded according to the steady-state density $\boldsymbol{\pi}$.  
Black arrows on the edges label the transition rates, and the colored edges indicate the steady-state current across each edge, $\boldsymbol{j}^\pi$.
}
\label{fig:graphfig}
\end{figure*}

Before studying the consequences of coarse-graining, we first build a detailed dynamical model.
At the smallest scales, the microscopic laws of physics are Markovian and deterministic.
By clustering microscopic configurations into sufficiently many mesostates, mesocopic dynamics may often be described by a continuous-time Markov jump process on the set of coarse-grained states~\cite{vankampen1992,Hoffmann2011,Esposito2012}.
Such a model can be constructed, for example, from atomistic simulations, as in the case of protein conformational dynamics~\cite{Chodera2007,Bowman2009}.
The jump process can be viewed as a random walk on a graph, with each of the $N$ vertices corresponding to one of the possible mesostates.
Fig.~\ref{fig:graphfig} depicts an example with twelve states.
Any two vertices, $y$ and $z$, are connected by an edge when the system can jump between $y$ and $z$.
We denote the rate for transitioning from $y$ to $z$ by $r(y,z)$ and require a non-vanishing rate $r(z,y)$ for the reverse transition.

The probability of occupying state $y$ at time $t$, $p_t(y)$, evolves according to the master equation
\begin{equation}
\frac{\partial p_t(y)}{\partial t} = - \sum_{z \neq y} j^{p_t}(y,z),
\end{equation}
where $j^{p}(y,z) = p(y) r(y, z) - p(z) r(z,y)$ is the current passing from $y$ to $z$.
Following Bertini et al.~\cite{Bertini2015Large,Bertini2015Flows}, the superscript $p$ highlights that this current is a function of the density.
At long times, the probability of occupying mesostate $y$ approaches the steady-state value,
\begin{equation}
\pi(y) \equiv \lim_{t\to\infty} p_t(y).
\end{equation}
Similarly, the current passing from state $y$ to $z$ approaches $j^\pi(y,z) = \pi(y) r(y,z) - \pi(z) r(z,y)$ in the steady state.
We will use bold symbols to represent vectors and let context distinguish whether the vector is indexed over vertices (as in $\boldsymbol{\pi}$) or edges (as in $\boldsymbol{j}^{\pi}$ and $\boldsymbol{r}).$

An equilibrium system obeys detailed balance, meaning that all steady-state currents vanish, $\boldsymbol{j}^{\pi} = \boldsymbol{0}$.
Non-vanishing currents may be generated if the system couples to multiple external reservoirs with differing intensive parameters, e.g., temperatures, pressures, or chemical potentials.
The environment, which we model as a set of infinitely-large reservoirs, can therefore lead the dynamics to break detailed balance, i.e., it need not be the case that mesostates $y$ and $z$ satisfy $\pi(y) r(y,z) = \pi(z) r(z,y).$

Because the mesoscopic Markov dynamics does not explicitly model the state of the thermodynamic reservoirs (how much energy and how many particles are in the baths), the impact of the reservoirs appears implicitly in the rates $\boldsymbol{r}$.
As an example, suppose mesostate $y$ has a (free) energy lower than that of mesostate $z$ by an amount $\Delta E$.
The transfer of energy from a thermal reservoir at inverse temperature $\beta$ can induce an uphill transition from $y$ to $z$.
If we assume the mesoscopic configurations are locally equilibrated with the thermal reservoir, then the transitions must obey detailed balance with respect to the reservoir's equilibrium distribution $p_{\rm eq}$~\cite{Esposito2010}.
This local detailed balance condition, $p_{\rm eq}(y) r(y,z) = p_{\rm eq}(z) r(z,y)$, yields
\begin{equation}
\frac{r(y,z)}{r(z,y)} = \frac{p_{\rm eq}(z)}{p_{\rm eq}(y)} = \exp(-\beta \Delta E),
\end{equation}
which relates the ratio of transition rates to the energy flux from the reservoir.
These same arguments apply to other types of reservoirs, e.g., a particle reservoir at constant chemical potential.
The ratio of transition rates, more generally, is expressed in terms of the entropy change in the reservoir, $\Delta S$:
\begin{equation}
\frac{r(y,z)}{r(z,y)} = \exp(\Delta S).
\label{eq:ldb}
\end{equation}

Satisfying local detailed balance with the various reservoirs does not imply that the Markov dynamics obeys detailed balance.
When coupled to multiple reservoirs with incompatible equilibrium states, the Markov dynamics breaks detailed balance, but Eq.~\eqref{eq:ldb} still relates the mesoscopic rates to the flow of entropy from the reservoirs. 
For density $\boldsymbol{p}$, we quantify the broken detailed balance by
\begin{equation}
F^{p}(y,z) = \ln \frac{p(y) r(y,z)}{p(z) r(z,y)},
\end{equation}
which we call the thermodynamic force because it measures how much free energy the baths must provide in order for the system to transition from $y$ to $z$~\footnote{$F^{p}(y,z)$ is the difference between the entropy gain of the bath $\Delta S = \ln r(y,z) / r(z,y)$ and the gain in Shannon entropy of the system $\ln p(z) / p(y)$.}.
In the steady state, the rate of that free energy transfer is given by
\begin{equation}
\sigma^\pi(y,z) = j^\pi(y,z) F^\pi(y,z).
\end{equation}
We call this quantity the dissipation rate for the edge connecting $y$ and $z$.
The total dissipation rate for the system is computed by summing the dissipation rates for all edges:
\begin{equation}
\Sigma^\pi = \boldsymbol{j}^{\pi} \cdot \boldsymbol{F}^{\pi} \equiv \sum_{y < z} j^\pi(y,z) F^{\pi}(y,z).
\label{eq:dissipationrate}
\end{equation}
The sum runs over pairs of states with $y<z$ to avoid double counting.

Computing the dissipation rate via Eq.~\eqref{eq:dissipationrate} is all-but-impossible for complicated problems.
The computation requires that all possible mesostates be identified, that the rates for transitions between these mesostates be measured, and that the steady-state density be computed.
Even the simplest of these tasks, identifying the set of mesostates, is frequently impractical.
More commonly, it is only possible to monitor transitions between some coarse-grained macrostates.
With this limitation in mind, we set out to infer the dissipation rate on the basis of fluctuations in finite-time stochastic trajectories.
Our strategy merely provides a lower bound on the dissipation rate, but may be applied even when we only observe macrostate transitions, a point we return to in Section~\ref{sec:macrostate}.

\subsection{Fluctuations}
An infinitely long trajectory samples all configurations in proportion to the steady-state distribution, but a single finite-time trajectory has fluctuations.
Consider one realization of the jump process, initialized in the steady state and observed for a long but finite time $T$.
We let $x(t)$ denote the identity of the occupied mesostate at time $t$.
Given this trajectory, an unbiased estimate of the steady-state density at mesostate $y$ is found by measuring the fraction of time spent in $y$:
\begin{equation}
p(y) = \frac{1}{T} \int_0^T dt \ \delta_{x(t), y},
\label{eq:empiricaldensity}
\end{equation}
with Kronecker delta $\delta_{\alpha, \beta}$.
This time-averaged density $\boldsymbol{p}$ is the \emph{empirical density}.
Despite similar notation, the empirical density should not be confused with the instantaneous probability of occupying state $y$ at time $t$, which we have denoted $p_t(y)$.
The instantaneous density $p_t(y)$ depends explicitly on time $t$, whereas the empirical density $p(y)$ depends on the timescale $T$ over which the density was averaged.

Analogous to the empirical density, the \emph{empirical current} from $y$ to $z$ counts the rate of transitions from $y$ to $z$, less those from $z$ to $y$:
\begin{equation}
j(y,z) \equiv \frac{1}{T} \int_0^T dt \ \delta_{x(t^-), y} \delta_{x(t^+),z} - \delta_{x(t^-), z} \delta_{x(t^+),y}.
\label{eq:empiricalcurrent}
\end{equation}
The notation $x(t^{\pm})$ is shorthand for the configuration immediately before or immediately after time $t$.
Note that $j(y,z)$ differs from the density-dependent current $j^p(y,z)$.
The former reflects the number of transitions observed in a stochastic trajectory while the latter is the current that one would expect given the empirical density $\boldsymbol{p}$ and the transition rates $\boldsymbol{r}$.

The probability distribution for the empirical density and current reflects the fluctuations anticipated in finite-time experiments or simulations.
For large $T$, this distribution adopts the large deviation form
\begin{equation}
P(\boldsymbol{p}, \boldsymbol{j}) \asymp e^{-T I(\boldsymbol{p}, \boldsymbol{j})}
\end{equation}
with the joint rate function $I(\boldsymbol{p}, \boldsymbol{j})$ measuring, on an exponential scale, the chance of observing fluctuations away from steady-state density ${\boldsymbol \pi}$ and steady-state current $\boldsymbol{j}^{\boldsymbol \pi}$~\cite{Touchette2009}.
We use $\asymp$ to indicate asymptotic equivalence, meaning
\begin{equation}
-\frac{1}{T} \ln P(\boldsymbol{p}, \boldsymbol{j}) =  I(\boldsymbol{p}, \boldsymbol{j}) + \mathcal{O}\left(\frac{1}{T}\right).
\end{equation}
The rate function attains a minimum at the steady-state values ($I({\boldsymbol \pi}, \boldsymbol{j}^{\pi}) \equiv 0$) since the steady-state behavior dominates in the long-time limit.
The local curvature of $I$ around its minimum, the Hessian, reveals the variance of the empirical fluctuations away from the steady-state values.
Remarkably, the local curvature can be computed analytically because an explicit expression for $I(\boldsymbol{p}, \boldsymbol{j})$ is known~\cite{Kesidis1993,Maes2008,Bertini2015Large,Bertini2015Flows}.

\section{Coarse-graining in Time}
\label{sec:heuristic}

The Markov jump process has rich temporal correlations\textemdash after a short time, a system initialized in mesostate $y$ is more likely to be in a neighboring state $z$ than the steady-state probability $\pi(z)$ would predict.
Though these correlations add complexity to transient dynamics, they die out after some finite correlation time.
The empirical density and empirical current can be averaged over sufficiently long times $T$ such that the transient dynamics becomes unimportant.
In this way, $\boldsymbol{p}$ and $\boldsymbol{j}$ can be thought of as the result of temporal coarse-graining with a coarse-graining timescale exceeding natural correlation times.
In the limit of large $T$, the empirical fluctuations can therefore be mimicked by Poisson point processes, which lack all temporal correlations.
More specifically, the coarse-grained Markov jump process resembles the behavior of a collection of independent Poisson point processes, one per directed edge of the graph, which count the number of transition events for each edge in time $T$.
The large deviation structure of Poisson processes are particularly simple, providing a heuristic route to $I(\boldsymbol{p}, \boldsymbol{j})$ for a general nonequilibrium jump process.
Rigorous derivations of this result utilize a technique called Cram\'{e}r tilting~\cite{Maes2008}.
Rather than recapitulate that argument, we illustrate how the form of the jump process rate function originates from a Poisson point process.

\subsection{Flows}

\begin{figure}
\centering
\includegraphics[width=0.45\textwidth]{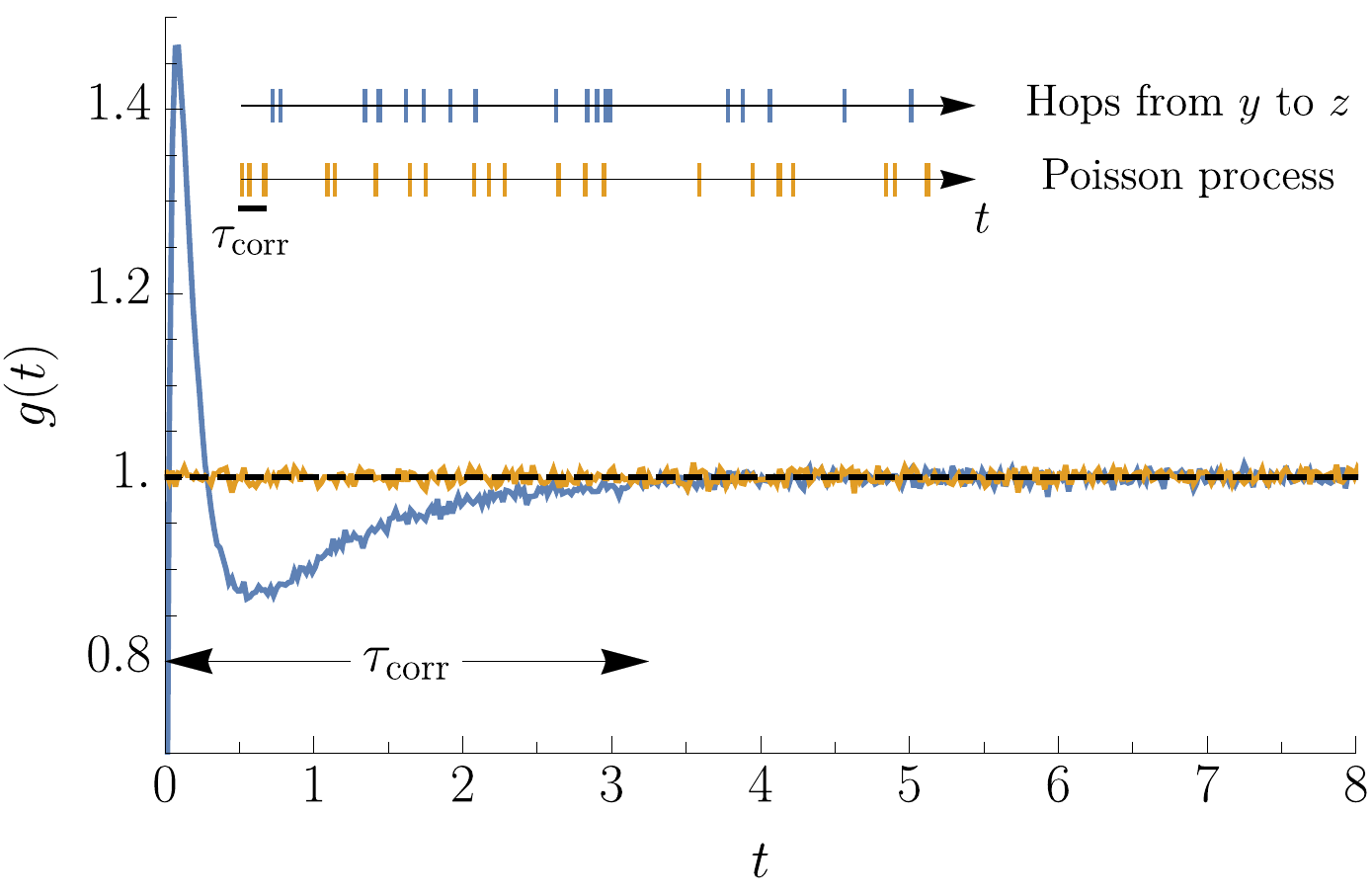}
\caption{Pair correlation functions $g(t)$ for Markov dynamics on the graph in Fig.~\ref{fig:graphfig} (blue) and for an Poisson point process (orange).
Time is reported in the same units as the inverse of the rate constants in Fig.~\ref{fig:graphfig}.
The colored lines are collected by sampling from 5000 trajectories, each of length 10000; the dashed black line shows the exact result for a Poisson point process.
The inset shows a timeline for the events observed in a representative trajectory, with each tick corresponding to the time of a hopping event.
The Poisson point process is constructed so that it has the same average density of tick marks as the Markov dynamics on the graph.
For times much larger than a correlation time $\tau_{\rm corr}$, the hops on the graph become uncorrelated, so the fluctuations in the number of events asymptotically approaches that of the Poisson point process, Eq.~\eqref{eq:poisson}.
} 
\label{fig:tickcorrelations}
\end{figure}

To explicitly demonstrate that the Markov jump process's long-time behavior resembles that of a collection of Poisson point processes, we first focus on a single type of dynamical event\textemdash hops from mesostate $y$ to $z$ in Fig.~\ref{fig:graphfig}.
We record the time of each such hop by a blue tick on a timeline in Fig.~\ref{fig:tickcorrelations}.
The average density of ticks is
\begin{equation}
q(y,z) = \frac{1}{T} \int_0^T dt \ \delta_{x(t^-), y} \delta_{x(t^+), z},
\end{equation}
the empirical flow rate from $y$ to $z$~\cite{Bertini2015Large, Bertini2015Flows}.
Due to the stochastic nature of the trajectories, the spacing between tick marks is variable.
Immediately following each $y \to z$ event, the system is localized in mesostate $z$, and the next $y \to z$ event cannot occur until the system finds its way back to mesostate $y$.
This need to reset introduces temporal correlations, quantified by the pair-correlation function
\begin{equation}
g(t) = \left<\frac{\sum_{i < j} \delta\left(t - \left(t_j - t_i\right)\right)}{q(y,z)}\right>,
\end{equation}
with $t_i$ denoting the time of the $i^{\rm th}$ $y \to z$ hop.
The average in the definition of $g(t)$ is taken over the ensemble of long trajectories of length $T$.

Akin to the pair correlation function of liquid state theory~\cite{Chandler1987}, $g(t)$ captures the probability that, given an event at time $0$, another event occurs after a delay time $t$.
The blue curve in Fig.~\ref{fig:tickcorrelations} shows $g(t)$ for the flows from state $y$ to $z$ in Fig.~\ref{fig:graphfig}.
At short times, $g(t)$ exceeds one, indicating a propensity for bursty repeated events since the return rate $r(z,y)$ can allow multiple $y \to z$ events in rapid succession.
After slightly longer times, $g(t)$ is less than one; trajectories which do not backtrack from $z$ to $y$ only enable the next event upon returning to $y$ via a slower, circuitous route.
Finally, $g(t)$ approaches $1$ for $t > \tau_{\rm corr}$, the timescale on which correlations are lost.

These temporal correlations are an important feature of a Markov jump process at short times.
However, the two-time correlations encoded in $g(t)$ become insignificant at long times, for which $g(t) \to 1$, resembling a Poisson point process~\footnote{The Poisson point process lacks correlations, so $g(t) = 1$.}.
This loss of correlations suggests that the probability of observing $Q$ transitions from $y$ to $z$ over a long time $T$ should be asymptotically given by a Poisson distribution~\cite{vankampen1992},
\begin{equation}
P(Q) = \frac{\lambda^Q e^{-\lambda}}{Q!}.
\label{eq:poissonwparameter}
\end{equation}
The Poisson parameter $\lambda$ must be chosen to match the steady-state flows in the long time limit, $\left<Q\right> = \pi(y) r(y,z) T$.
We note that this condition can be met by choosing $\lambda = T p(y) r(y,z)$ if we average over both empirical densities and flows.
This observation is suggestive of a long-time Poisson form with an effective rate that depends on the empirical density $p(y)$:
\begin{equation}
P_{\rm ind}(p(y), q(y,z)) \asymp \frac{(T p(y) r(y,z))^{T q(y,z)} e^{-T p(y) r(y,z)}}{(T q(y,z))!},
\label{eq:poisson}
\end{equation}
with the subscript ``ind'' denoting that we have considered the single edge as independent of the other edges.
In reality, the statistics of neighboring edges are coupled by a conservation law: a trajectory must leave one state to enter another, so the empirical flow conserves probability at every mesostate~\footnote{The initial and final mesostates do not conserve probability\textemdash there is a source where the system starts at time zero and a sink where it ends at time $T$, but in the long time limit this effect is insignificant.}, which requires
\begin{equation}
\sum_z \left(q(y,z) - q(z,y)\right)= 0 \ \forall y.
\label{eq:conservative}
\end{equation}
$P_{\rm ind}$ can be thought of as the effective single-edge distribution in the absence of the conservation law constraint.
We stress that $P_{\rm ind}$ should not be confused with the marginal distribution for the single-edge statistics, which we will see is much more complicated.

The joint statistics of density and flow across all edges, however, can be simply expressed.
This simplicity arises because the long-time flow fluctuations on the various edges are coupled only by the constraint Eq.~\eqref{eq:conservative}, which requires that the empirical flow $\boldsymbol{q}$ must conserve probability.
When the empirical flow is conservative, the long-time joint probability of $\boldsymbol{p}$ and $\boldsymbol{q}$ can be written as a product over the independent edge probabilities $P_{\rm ind}$.
Any vector $\boldsymbol{q}$ that does not satisfy the conservation law, of course, has vanishing probability in the long-time limit.
To write this claim in large deviation form we first rewrite $P_{\rm ind}$ in large deviation form, $P_{\rm ind}(p(y), q(y,z)) \asymp e^{-T I_{\rm ind}(p(y), q(y,z))}$, with the rate function
\begin{equation}
I_{\rm ind}(p(y), q(y,z)) = p(y) r(y,z) - q(y,z) + q(y,z) \ln \frac{q(y,z)}{p(y) r(y,z)}.
\label{eq:ratefunction}
\end{equation}
The joint density and flow fluctuations are then given by
\begin{equation}
I(\boldsymbol{p}, \boldsymbol{q}) = \begin{cases}
\sum_{y < z} I_{\rm ind}(p(y), q(y,z)), \ \boldsymbol{q} \text{ satisfies Eq.~\eqref{eq:conservative}}\\
\infty, \ \text{otherwise.}
\end{cases}
\label{eq:jointratefunction}
\end{equation}
This form of $I(\boldsymbol{p}, \boldsymbol{q})$ reveals why $P_{\rm ind}$ is not the marginal distribution for single-edge flow statistics.
The conservation law constraint complicates any attempts to simply express the marginal $P(p(y), q(y,z)$ by integrating out the other densities and flows.

Our heuristic arguments for the form of $I(\boldsymbol{p}, \boldsymbol{q})$ are suggestive, but they do not constitute a proof so much as a motivation.
Eq.~\eqref{eq:jointratefunction}, known in mathematics literature as the Level 2.5 large deviation function, can be proven using more sophisticated arguments that construct an effective process to generate rare densities and flows~\cite{Maes2008, Bertini2015Flows}.
While that construction offers rigor and alternative insight, we find our simple, heuristic derivation to be instructive since it clearly identifies the essential physics: we may discard temporal correlations in a complicated Markov jump process to obtain an asymptotically equivalent collection of Poisson point processes.

\subsection{Currents}
Suppose now that we are interested in the empirical currents rather than the flows.
Unlike $q(y,z)$, $j(y,z)$ deducts the rate of reversed hops from $z$ to $y$.
The probability of measuring a current $j(y,z)$ is thus given by marginalizing over $q(y,z)$ and $q(z,y)$ with the constraint $j(y,z) = q(y,z) - q(z,y)$.
As with the flows, currents across different edges are coupled by probability conservation:
\begin{equation}
\sum_z j(y,z) = 0  \ \forall y.
\label{eq:conservationcurrents}
\end{equation}
For large $T$, this marginalization of flows proceeds via a saddle point approximation~\cite{Touchette2009} to yield $P(\boldsymbol{p}, \boldsymbol{j}) \asymp e^{-T I(\boldsymbol{p}, \boldsymbol{j})}$ with
\begin{equation}
I(\boldsymbol{p}, \boldsymbol{j}) = \begin{cases}
\sum_{y<z} \Psi(p(y), p(z), j(y,z)), \ \boldsymbol{j} \text{ satisfies Eq.~\eqref{eq:conservationcurrents}}\\
\infty, \ \text{otherwise.}
\end{cases}
\label{eq:jumpratefunc}
\end{equation}
The function $\Psi$ follows from treating each edge independently.
For example, integrating out $q(y,z)$ and $q(z,y)$ yields
\begin{align}
\nonumber \Psi(&p(y), p(z), j(y,z)) \\
&=\inf_{q(y,z)} I\Big(p(y), q(y,z)\Big)  + I\Big(p(z), q(y,z) - j(y,z)\Big).
\end{align}
The minimizer $q_\star(y,z)$ is the root of a quadratic,
\begin{equation}
q_\star(y,z) = \frac{1}{2} \left(j(y,z) + \sqrt{j(y,z)^2 + a^p(y,z)^2}\right),
\label{eq:qstar}
\end{equation}
where $a^p(y,z) \equiv 2 \sqrt{p(y)p(z)r(y,z)r(z,y)}$.
Thus $\Psi$ may be expressed in terms of $q_\star$ as
\begin{align}
\nonumber \Psi(&p(y), p(z), j(y,z))\\
\nonumber &= I\big(p(y), q_\star(y,z)\big) + I\big(p(z), q_\star(z,y)\big)\\
\nonumber &= \Bigg [p(y) r(y,z) + p(z) r(z,y) - q_\star(y,z) - q_\star(z,y)\\ 
& \ \ \ \ \ \ + q_\star(y,z) \ln \frac{q_\star(y,z)}{p(y) r(y,z)} + q_\star(z,y)\ln \frac{q_\star(z,y)}{p(z) r(z,y)}\bigg].
\label{eq:psi}
\end{align}
In Appendix~\ref{app:ratefnderivation} we carry out straightforward algebraic manipulations to bring $\Psi$ into the form of Bertini et al.~\cite{Bertini2015Flows},
\begin{align}
\nonumber \Psi &= \sqrt{j^p(y,z)^2 + a^p(y,z)^2} - \sqrt{j(y,z)^2 + a^p(y,z)^2}\\
& \ \ + j(y,z) \left(\arcsinh \frac{j(y,z)}{a^p(y,z)} - \arcsinh \frac{j^p(y,z)}{a^p(y,z)}\right).
\label{eq:psibertini}
\end{align}

\section{Rate function bound}
\label{sec:ratefunctionbound}
\begin{figure}
\centering
\includegraphics[width=0.45\textwidth]{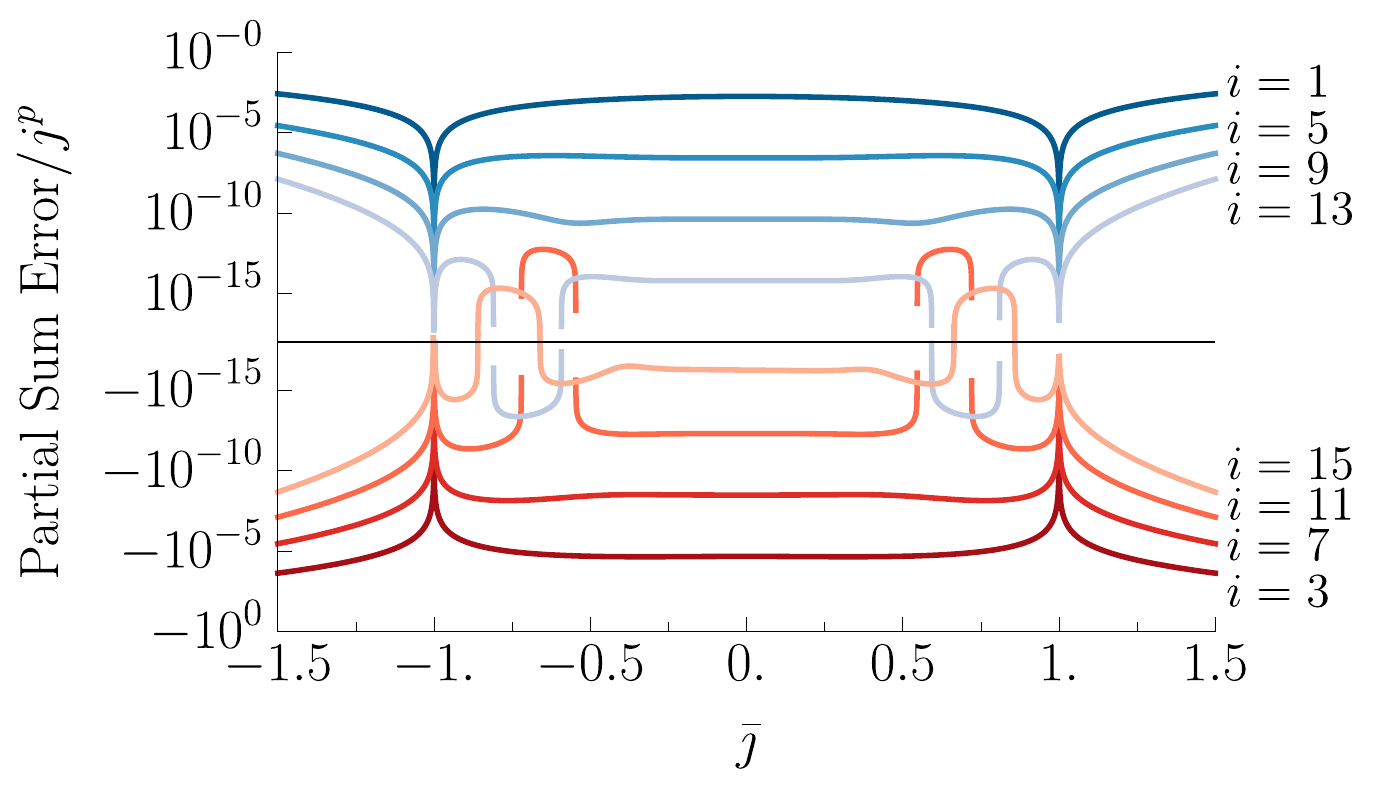}
\caption{Residuals from approximating $\Psi$ by the partial sums of the Taylor series expansion in $F^p$, Eq.~\eqref{eq:Taylor}, plotted for $F^p = 0.75$.
The label $i$ indicates that the partial sum is computed up through (and including) the $(F^p)^i$ term in the series.
The residual is this $i^{\rm th}$ partial sum minus $\Psi$.
When $\bar{\jmath}$ is in the neighborhood of $\pm 1$, the Taylor expansion for $\Psi$ is an alternating series, so the partial sums provide upper and lower bounds on $\Psi$.
For low-order partial sums and for small $F^p$ the partial sums provide bounds for all values of $\bar{\jmath}$.
As an example, the $i = 1$ partial sum is the quadratic bound given by Eq.~\eqref{eq:quadraticbound}, which holds for all values of $\bar{\jmath}$ and $F^p$.
}
\label{fig:partialsums}
\end{figure}

To relate the fluctuations described by the rate function $I(\boldsymbol{p}, \boldsymbol{j})$ to the thermodynamic dissipation, it is useful to write $I$ in terms of the currents and thermodynamic forces $\boldsymbol{F}^p$.
For notational compactness we suppress the labels of the vertices.
We note that $j^p / \sinh(F^p/2) = a^p \geq 0$, implying that $j^p$ has the same sign as $F^p$.
Furthermore, it is natural to measure the empirical current $j$ relative to the current $j^p$, so we introduce $\bar{\jmath} \equiv j / j^p$.
With this new notation,
\begin{align}
\nonumber \Psi &= j^p \bigg [ \coth\frac{F^p}{2} -\frac{\bar{\jmath} F^p}{2} + \bar{\jmath} \arcsinh\left(\bar{\jmath} \sinh\frac{F^p}{2}\right)\\
& \ \ \ \ \ \ \ - \sqrt{\bar{\jmath}^2 + \csch^2\frac{F^p}{2}}\bigg].
\end{align}
Taylor expanding in powers of $F^p$, we obtain
\begin{align}
\nonumber \Psi &= j^p \bigg[\frac{(\bar{\jmath} - 1)^2 F^p}{4} - \frac{(\bar{\jmath}^2 - 1)^2 (F^p)^3}{192} \\
& \ \ \ \ \ \ \ \ \ \ \ \ \ + \frac{(\bar{\jmath}^2 - 1)^2(3\bar{\jmath}^2 + 1) (F^p)^5}{7680} + \mathcal{O}((F^p)^7)\bigg].
\label{eq:Taylor}
\end{align}
The low-order partial sums of this series alternate, first overestimating, then underestimating $\Psi$ (see Fig.~\ref{fig:partialsums}).
In particular, the first partial sum bounds the rate function by a quadratic~\cite{Gingrich2016},
\begin{equation}
\Psi \leq \Psi_{\rm quad} \equiv \frac{(\bar{\jmath} - 1)^2 \sigma^p}{4},
\label{eq:quadraticbound}
\end{equation}
where $\sigma^p = j^p F^p$ is the local dissipation for the edge.

Truncation of a Taylor series, of course, does not necessarily yield a bound.
To prove inequality~\eqref{eq:quadraticbound} we must confirm the positivity of the residual $\Delta \equiv \Psi_{\rm quad} - \Psi$.
Since $\Delta$ is symmetric about $\bar{\jmath} = 0$, it suffices to consider positive $\bar{\jmath}$.
We note that $\Delta$ vanishes when $\bar{\jmath} = 1$, and 
\begin{equation}
\frac{\partial \Delta}{\partial \bar{\jmath}} = j^p \left[\arcsinh\left(\bar{\jmath}\sinh\frac{F^p}{2}\right) - \frac{\bar{\jmath} F^p}{2}\right],
\label{eq:deriv}
\end{equation}
implying that $\Delta$ strictly increases as $|\bar{\jmath}-1|$ grows~\footnote{Using the monotonicity of $\arcsinh$ and the concavity of $\sinh$, Eq.~\eqref{eq:deriv} implies that $\partial \Delta / \partial \bar{\jmath} > 0$ for $\bar{\jmath} > 1$ and $\partial \Delta / \partial \bar{\jmath} < 0$ for $0 \leq \bar{\jmath} \leq 1$.}.
Since $\Psi \leq \Psi_{\rm quad}$ is bounded for every edge of the graph, $I(\boldsymbol{p}, \boldsymbol{j})$ is bounded by a quadratic form whose curvature is determined by the local dissipation rates:
\begin{equation}
I(\boldsymbol{p}, \boldsymbol{j}) \leq I_{\rm quad}(\boldsymbol{p}, \boldsymbol{j}) \equiv \sum_{y<z} \frac{\sigma^p(y,z)}{4 j^p(y,z)^2} \left(j(y,z) - j^p(y,z)\right)^2,
\label{eq:quadbound}
\end{equation}
Eq.~\eqref{eq:quadbound} is restricted to conservative currents, or alternatively we take $I_{\rm quad} \equiv \infty$ for nonconservative $\boldsymbol{j}$.

Recall that by construction $I(\boldsymbol{\pi}, \boldsymbol{j}^{\pi}) = 0$.
Our quadratic bound shares this minimum.
Furthermore, $I(\boldsymbol{\pi}, -\boldsymbol{j}^{\pi}) = I_{\rm quad}(\boldsymbol{\pi}, -\boldsymbol{j}^{\pi})$, so $I_{\rm quad}$ is the tightest quadratic upper bound which has a minimum at $(\boldsymbol{\pi}, \boldsymbol{j}^{\pi})$ and can be expressed as a single sum over edges (i.e., is diagonal in the edge basis)~\cite{Polettini2016}.
The curvature of $I$, evaluated at $\boldsymbol{\pi}$ and $\boldsymbol{j}^{\pi}$, reflects the variance of long-time current fluctuations.
The tighter inequality ~\eqref{eq:quadbound} is, the more precisely we may relate this variance to the local dissipation rates.
Truncation after the first-order term of the series expansion Eq.~\eqref{eq:Taylor} becomes exact for small thermodynamic forces, meaning that the quadratic bound is accurate when $F^p$ is small.

\section{Scalar Current Fluctuations}
\label{sec:scalarcurrent}
The large deviation function $I(\boldsymbol{p}, \boldsymbol{j})$ describes the joint distribution for the density in all mesostates and currents between these states.
Experiments, however, cannot hope to resolve the statistical details of a large number of degrees of freedom.
To make practical use of inequality~\eqref{eq:quadbound}, we must project the result onto a smaller probability space.
Rather than monitoring $\boldsymbol{p}$ and $\boldsymbol{j}$, suppose we only measure a scalar current of the form
\begin{equation}
j_d = \boldsymbol{j} \cdot \boldsymbol{d} \equiv \sum_{y<z} j(y,z) d(y,z).
\label{eq:jddef}
\end{equation}
The generalized current $j_d$ is a linear combination of the currents between mesostates with expansion coefficients $\boldsymbol{d}$~\cite{Barato2015}.
This construction affords significant flexibility.
By choosing $\boldsymbol{d} = \boldsymbol{F}^{\pi}$, the generalized current is the same as the dissipation rate.
Other choices of $\boldsymbol{d}$ can highlight the current across a single edge or the current associated with transitions between macrostates, as illustrated in Section~\ref{sec:macrostate}.

In the long-time limit, the distribution for this generalized current adopts the large deviation form $P(j_d) \asymp e^{-T I(j_d)}$.
Because $P(j_d)$ can be constructed from $P(\boldsymbol{p}, \boldsymbol{j})$, the rate function for the generalized current is related to $I(\boldsymbol{p}, \boldsymbol{j})$ ~\footnote{The restriction to conservative currents is built into $I$ because $I(\boldsymbol{p}, \boldsymbol{j}) = \infty$ for non-conservative $\boldsymbol{j}$}:
\begin{equation}
I(j_d) = \inf_{\boldsymbol{p}, \boldsymbol{j} | \boldsymbol{j} \cdot \boldsymbol{d} = j_d} I(\boldsymbol{p}, \boldsymbol{j}).
\label{eq:IjdAsinf}
\end{equation}
This infimum is bounded from above by $I(\boldsymbol{p}^*, \boldsymbol{j}^*)$ for any choice of $\boldsymbol{p}^*$ and conservative $\boldsymbol{j}^*$ such that $\boldsymbol{j}^* \cdot \boldsymbol{d} = j_d$.
We choose $\boldsymbol{p}^* = \boldsymbol{\pi}$ and $\boldsymbol{j}^* = (j_d / j_d^\pi) \boldsymbol{j}^{\pi}$, where $j_d^\pi \equiv \boldsymbol{j}^{\pi} \cdot \boldsymbol{d}$.
As a multiple of the conservative steady-state current $\boldsymbol{j}^{\pi}$, $\boldsymbol{j}^*$ is guaranteed to be conservative.
Hence
\begin{align}
\nonumber I(j_d) &\leq I(\boldsymbol{p}^*, \boldsymbol{j}^*)\\
\nonumber &= \frac{1}{4}\left(\frac{j_d}{j_d^\pi} - 1\right)^2 \sum_{y<z} \sigma^\pi(y,z)\\
&= \frac{(j_d - j_d^\pi)^2 \Sigma^\pi}{4(j_d^\pi)^2},
\label{eq:ratefunctionbound}
\end{align}
where we have used inequality~\eqref{eq:quadbound}.
The bound on the large deviation function translates to a bound on the variance of $j_d$ since $\text{var}(j_d) = 1 / I''(j_d^\pi)$, so measuring the mean and variance of any scalar current provides a lower bound on the dissipation rate:
\begin{equation}
\frac{2 (j_d^\pi)^2}{\text{var}(j_d)} \leq \Sigma^\pi.
\label{eq:smalldevbound}
\end{equation}

The bound on $\Sigma^\pi$ is most useful if it is tight, but there are two distinct reasons it might be loose: (1) $\Psi$ for the edges could deviate significantly from the quadratic bound $\Psi_{\rm quad}$ or (2) our choice of $\boldsymbol{p}^*$ and $\boldsymbol{j}^*$ could be suboptimal.
In the remainder of the paper we study diffusion processes as a limit of Markov jump processes.
That limiting procedure yields $\Psi = \Psi_{\rm quad}$, implying that the diffusion process bound is weakened only by our suboptimal $\boldsymbol{p}^*$ and $\boldsymbol{j}^*$.
For diffusions, the high-dimensional rate function $I(\boldsymbol{p}, \boldsymbol{j})$ is exactly given by the quadratic $I_{\rm quad}(\boldsymbol{p}, \boldsymbol{j})$, but the low-dimensional rate function $I(j_d)$ can still differ significantly from the upper bound in inequality~\eqref{eq:ratefunctionbound}.
Section~\ref{sec:macrostate} considers this scenario in greater detail.

\section{Diffusion processes}
\label{sec:diffusion}

\begin{figure}[t!]
\centering
\includegraphics[width=0.45\textwidth]{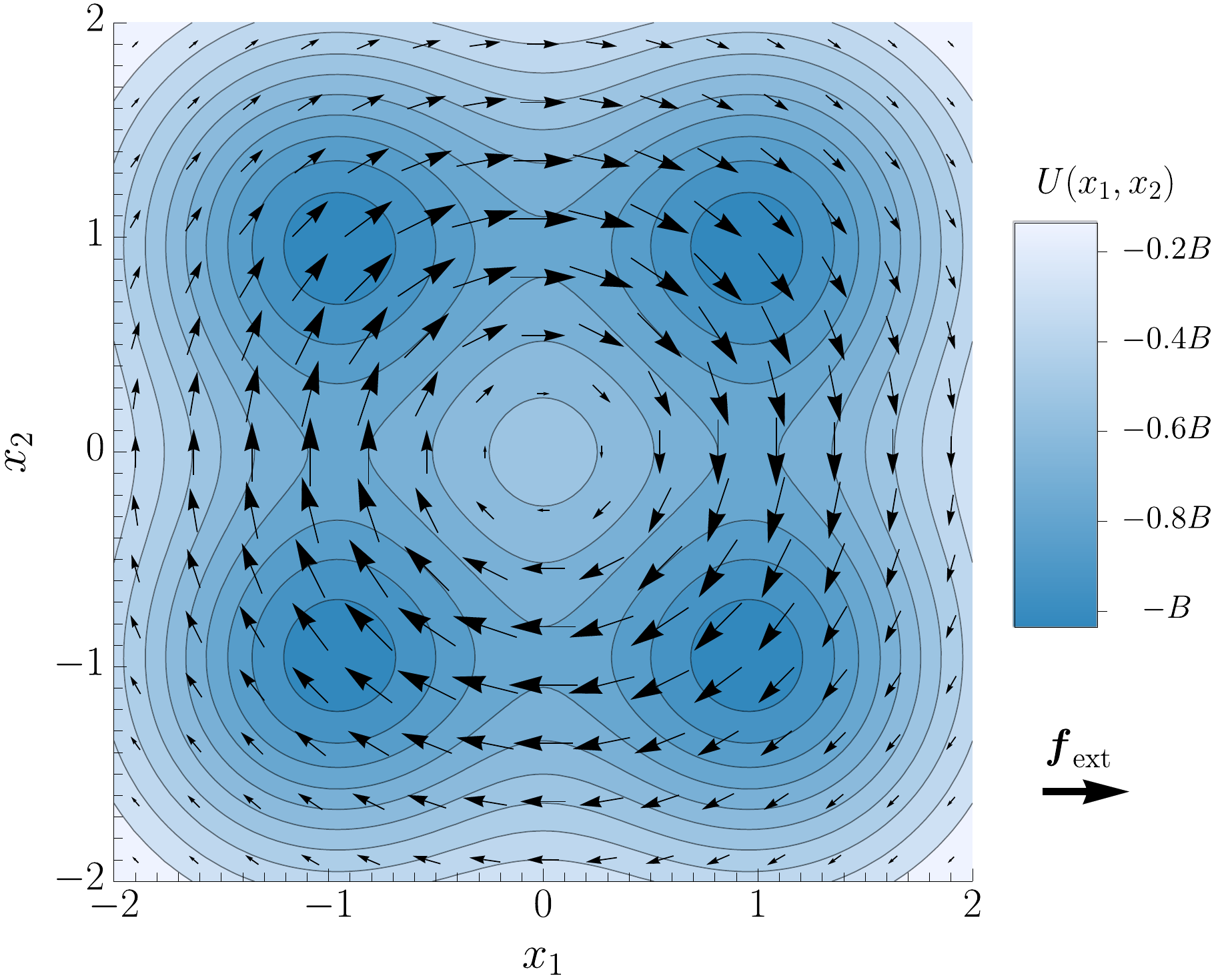}
\caption{Contour plot of a free energy $U(\boldsymbol{x}) = - B \sum_{i=1}^4 e^{-(\boldsymbol{x} - \boldsymbol{c}_i)^2}$, with the $\boldsymbol{c}_i$'s setting the center of the Gaussian wells at $(\pm 1, \pm 1)$.
The parameter $B$ controls the barrier heights (or equivalently the well depths).
The black arrows represent a vector field of the non-gradient external field, $\boldsymbol{f}_{\rm ext}(\boldsymbol{x}) = A \boldsymbol{x}^2 e^{-3|\boldsymbol{x}|}(x_2, -x_1)$, which drives cycles around the origin.
The external field's amplitude is regulated by the parameter $A$.
}
\label{fig:diffusionfig}
\end{figure}

Though our results have been derived for Markov jump processes, we may translate them into appropriate forms to describe fluctuations in diffusion processes.
For simplicity, we consider this diffusive limit only for a single particle moving in two dimensions with mobility $\mu$. 
Let $\boldsymbol{x} \equiv (x_1, x_2)$ denote the particle's position, which evolves according to an overdamped Langevin equation with deterministic force $\boldsymbol{f}(\boldsymbol{x}) \equiv (f_1(\boldsymbol{x}), f_2(\boldsymbol{x}))$ and random force $\boldsymbol{\eta}$.
We further decompose the deterministic force into a contribution from a free energy gradient $\nabla U(\boldsymbol{x})$ and one from a non-gradient external field $\boldsymbol{f}_{\rm ext}(\boldsymbol{x})$,
\begin{equation}
\frac{\partial \boldsymbol{x}}{\partial t} = - \mu \nabla U(\boldsymbol{x}) + \mu \boldsymbol{f}_{\rm ext}(\boldsymbol{x})  + \boldsymbol{\eta}.
\label{eq:langevin}
\end{equation}
The random force at time $t$, $\boldsymbol{\eta}(t) \equiv (\eta_1(t), \eta_2(t))$ is a vector of Gaussian random variables satisfying $\left<\eta_k(t)\right> = 0$ and $\left<\eta_k(t)\eta_{k'}(t')\right> = 2 D\delta_{k, k'} \delta(t - t')$, where $D$ is the diffusion constant.
In Section~\ref{sec:macrostate} we will consider the particular driven diffusive process shown in Fig.~\ref{fig:diffusionfig} as an illustrative example.

Associated to the Langevin equation is a Fokker-Planck equation describing the evolution of probability density at position $\boldsymbol{x}$, $\rho(\boldsymbol{x})$:
\begin{equation}
\frac{\partial \rho(\boldsymbol{x})}{\partial t} = -\nabla \cdot \boldsymbol{\mathcal{J}}^\rho(\boldsymbol{x}),
\label{eq:fokkerplanck}
\end{equation}
where 
\begin{equation}
\boldsymbol{\mathcal{J}}^\rho(\boldsymbol{x}) \equiv (\mathcal{J}_1^\rho(\boldsymbol{x}), \mathcal{J}_2^\rho(\boldsymbol{x})) = \boldsymbol{f}(\boldsymbol{x}) \rho(\boldsymbol{x}) - D \nabla \rho(\boldsymbol{x})
\end{equation}
is the $\rho$-dependent current at $\boldsymbol{x}$.
We denote the steady-state density $\rho^\pi$ and the steady-state current $\boldsymbol{\mathcal{J}}^\pi$.
As in the jump process, we define an empirical density field and empirical current field, $\rho$ and $\boldsymbol{\mathcal{J}}$ respectively, and measure the probability of fluctuations away from $\rho^\pi$ and $\boldsymbol{\mathcal{J}}^\pi$ using the rate function $\mathcal{I}[\rho, \boldsymbol{\mathcal{J}}]$.
The square brackets highlight that $\mathcal{I}$ is now a functional of the density and current fields.
The exact form of $\mathcal{I}$ is known~\cite{Barato2015Formal, Polettini2016}.
In this section, we give a complementary derivation of the result using a limit of the Markov jump process results.
The limiting procedure serves to clarify the origin of quadratic current fluctuations in diffusion processes.

\begin{figure*}[th]
\centering
\includegraphics[width=0.95\textwidth]{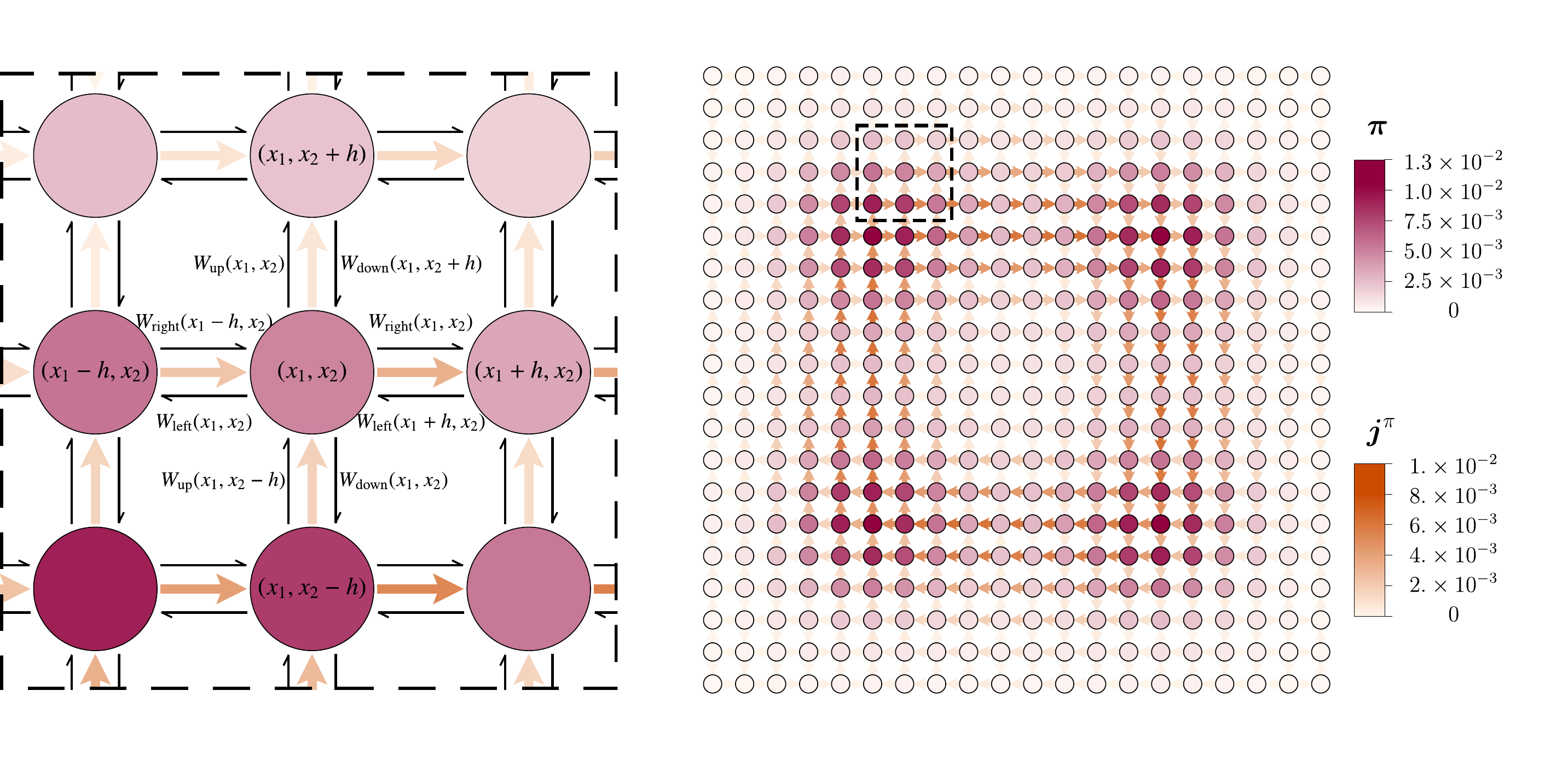}
\caption{A jump process approximates the diffusion in Fig.~\ref{fig:diffusionfig}.
On the left, a close-up view of a sub-system shows the transition rates on the lattice.
On the right, the steady-state behavior of the full state space $(\boldsymbol{x} \in [-2,2] \times [-2,2])$ is shown for $A = 8, B = 5, D = 1, \mu = 1$.
The steady-state density primarily resides in the four wells, but the external field drives some clockwise current around the origin.
}
\label{fig:discretizationfig}
\end{figure*}

To leverage our previous results, we approximate the diffusion process by a jump process on a square lattice, as depicted in Fig.~\ref{fig:discretizationfig}.
There is no unique way to coarse-grain a diffusion to a jump process~\cite{Gardiner2009}.
Thus we have freedom in how we construct our model so long as it yields a diffusive limit that matches the Fokker-Planck equation, Eq.~\eqref{eq:fokkerplanck}, when the lattice spacing becomes infinitesimal.
We construct a simple nearest-neighbor jump process which is entirely characterized by four space-dependent transition rates for hopping from a grid point at $\boldsymbol{x}$ to a nearest-neighbor site: $W_{\rm up}(\boldsymbol{x}), W_{\rm right}(\boldsymbol{x}), W_{\rm down}(\boldsymbol{x}),$ and $W_{\rm left}(\boldsymbol{x})$.
These rates must scale with the lattice spacing $h$ in such a way that the first two jump moments give the correct drift and diffusion~\cite{Gardiner2009}, requiring
\begin{align}
\nonumber h(W_{\rm right}(\boldsymbol{x}) - W_{\rm left}(\boldsymbol{x})) &= \mu f_1(\boldsymbol{x})\\
\nonumber h(W_{\rm up}(\boldsymbol{x}) - W_{\rm down}(\boldsymbol{x})) &= \mu f_2(\boldsymbol{x}) \\
\nonumber h^2(W_{\rm right}(\boldsymbol{x}) + W_{\rm left}(\boldsymbol{x})) &= 2D\\
h^2(W_{\rm up}(\boldsymbol{x}) + W_{\rm down}(\boldsymbol{x})) &= 2D,
\end{align}
as $h \to 0$. 
From these constraints the hopping rates are
\begin{align}
\nonumber W_{\rm right}(\boldsymbol{x}) &= \frac{\mu f_1(\boldsymbol{x})}{2h} + \frac{D}{h^2}\\
\nonumber W_{\rm left}(\boldsymbol{x}) &= \frac{-\mu f_1(\boldsymbol{x})}{2h} + \frac{D}{h^2}\\
\nonumber W_{\rm up}(\boldsymbol{x}) &= \frac{\mu f_2(\boldsymbol{x})}{2h} + \frac{D}{h^2}\\
W_{\rm down}(\boldsymbol{x}) &= \frac{-\mu f_2(\boldsymbol{x})}{2h} + \frac{D}{h^2}.
\label{eq:discretizationrates}
\end{align}
We now identify each vertex of the graph by its location $\boldsymbol{x}$, writing the density at that vertex as $p(\boldsymbol{x})$.
The edge connecting grid point $\boldsymbol{x}$ to its neighbor on the right has current $j^p_{\rm right}(\boldsymbol{x}) \equiv j^p(\boldsymbol{x}, \boldsymbol{x} + (h,0))$, and the thermodynamic force associated to that edge is $F^p_{\rm right}(\boldsymbol{x}) \equiv F^p(\boldsymbol{x}, \boldsymbol{x} + (h,0))$.
The ``up,'' ``down,'' and ``left'' directions are treated analogously.
Using this notation, we may rewrite $I_{\rm quad}$ of Eq.~\eqref{eq:quadbound} as
\begin{align}
\nonumber I_{\rm quad}(\boldsymbol{p}, \boldsymbol{j}) = \frac{1}{8} &\sum_{\boldsymbol{x}} \bigg[\frac{F^p_{\rm left}(\boldsymbol{x})}{j^p_{\rm left}(\boldsymbol{x})} (j_{\rm left}(\boldsymbol{x}) - j^p_{\rm left}(\boldsymbol{x}))^2\\
\nonumber &+\frac{F^p_{\rm right}(\boldsymbol{x})}{j^p_{\rm right}(\boldsymbol{x})} (j_{\rm right}(\boldsymbol{x}) - j^p_{\rm right}(\boldsymbol{x}))^2\\
\nonumber &+\frac{F^p_{\rm down}(\boldsymbol{x})}{j^p_{\rm down}(\boldsymbol{x})} (j_{\rm down}(\boldsymbol{x}) - j^p_{\rm down}(\boldsymbol{x}))^2\\
&+\frac{F^p_{\rm up}(\boldsymbol{x})}{j^p_{\rm up}(\boldsymbol{x})} (j_{\rm up}(\boldsymbol{x}) - j^p_{\rm up}(\boldsymbol{x}))^2\bigg].
\label{eq:Iquadlong}
\end{align}

To simplify the expression further, we must convert from the discrete-space density $\boldsymbol{p}$ and current $\boldsymbol{j}$ to the fields $\rho$ and $\boldsymbol{\mathcal{J}}$.
In the continuum limit, $\boldsymbol{p}(\boldsymbol{x})$ vanishes while $\rho(\boldsymbol{x})$ remains finite such that $\boldsymbol{p}(\boldsymbol{x}) \to h^2 \rho(\boldsymbol{x})$.
The current field $\boldsymbol{\mathcal{J}}^\rho$ is also finite, but the current on any edge of the lattice vanishes with order $h$:
\begin{align}
\nonumber j^p_{\rm right}(\boldsymbol{x}) &= p(\boldsymbol{x}) W_{\rm right}(\boldsymbol{x}) - p\Big(\boldsymbol{x} + (h,0)\Big) W_{\rm left}\Big(\boldsymbol{x} + (h,0)\Big)\\
\nonumber &=\left(\mu f_1(\boldsymbol{x}) \rho(\boldsymbol{x}) - D \frac{\partial \rho(\boldsymbol{x})}{\partial x_1}\right)h + \mathcal{O}(h^2)\\
&=\mathcal{J}^\rho_1(\boldsymbol{x}) h + \mathcal{O}(h^2).
\label{eq:currentsmallh}
\end{align}
Likewise, the thermodynamic force on each edge vanishes in proportion to $h$:
\begin{align}
\nonumber F^p_{\rm right}(\boldsymbol{x}) &= \ln \frac{p(\boldsymbol{x}) W_{\rm right}(\boldsymbol{x})}{p\Big(\boldsymbol{x} + (h,0)\Big) W_{\rm left}\Big(\boldsymbol{x} + (h,0)\Big)}\\
\nonumber &= \left(\frac{\mu f_1(\boldsymbol{x})}{D} - \frac{1}{p(\boldsymbol{x})} \frac{\partial p(\boldsymbol{x})}{\partial x_1}\right)h + \mathcal{O}(h^2)\\
&= \mathcal{F}_1^\rho(\boldsymbol{x})h + \mathcal{O}(h^2),
\label{eq:forcesmallh}
\end{align}
with a finite thermodynamic force field given by
\begin{equation}
\boldsymbol{\mathcal{F}}^\rho(\boldsymbol{x}) \equiv (\mathcal{F}^\rho_1(\boldsymbol{x}), \mathcal{F}^\rho_2(\boldsymbol{x})) = \frac{\mu \boldsymbol{f}(\boldsymbol{x})}{D} - \nabla \ln \rho(\boldsymbol{x}).
\end{equation}
Hence the ratio of force to current on an edge, which appears in each term of Eq.~\eqref{eq:Iquadlong}, remains finite and independent of the jump direction,
\begin{equation}
\lim_{h \to 0} \frac{F^p_{\rm right}(\boldsymbol{x})}{j^p_{\rm right}(\boldsymbol{x})} = \frac{\mathcal{F}_1^\rho(\boldsymbol{x})}{\mathcal{J}_1^\rho(\boldsymbol{x})} = \frac{\frac{\mu f_1}{D} - \frac{\partial}{\partial x_1} \ln \rho(\boldsymbol{x})}{\mu f_1\rho - D \frac{\partial}{\partial x_1} \rho(\boldsymbol{x})} = \frac{1}{D \rho(\boldsymbol{x})}.
\label{eq:Fjratio}
\end{equation}

Recall from Eq.~\eqref{eq:Taylor} that the rate function bound~\eqref{eq:quadbound} becomes tighter for small thermodynamic forces.
Since the forces vanish in the $h \to 0$ limit, the jump process on an infinitesimal grid thus saturates inequality~\eqref{eq:quadbound}, yielding the equality
\begin{equation}
\mathcal{I}[\rho, \boldsymbol{\mathcal{J}}] = \lim_{h \to 0} I(\boldsymbol{p}, \boldsymbol{j}) = \lim_{h \to 0} I_{\rm quad}(\boldsymbol{p}, \boldsymbol{j}).
\end{equation}
In other words, approximating $I$ by $I_{\rm quad}$ becomes exact in the continuum limit.
We convert the sum to an integral using
\begin{equation}
\sum_{\boldsymbol{x}} h^2 \to \int d\boldsymbol{x}
\end{equation}
and insert Eqs.~\eqref{eq:currentsmallh} and~\eqref{eq:Fjratio} into Eq.~\eqref{eq:Iquadlong} to obtain the rate function for diffusions
\begin{equation}
\mathcal{I}[\rho, \boldsymbol{\mathcal{J}}] = \int d\boldsymbol{x} \frac{(\boldsymbol{\mathcal{J}}(\boldsymbol{x}) - \boldsymbol{\mathcal{J}}^\rho(\boldsymbol{x}))^2}{4 D \rho(\boldsymbol{x})}.
\end{equation}
Note that the current fluctuations are strictly quadratic, a fact that can be traced back to the vanishingly small thermodynamic force on each infinitesimal edge of the graph.

The rate function may alternatively be expressed with respect to the local dissipation rates,
\begin{equation}
\boldsymbol{\sigma}^\rho(\boldsymbol{x}) = \boldsymbol{\mathcal{J}}^\rho(\boldsymbol{x}) \cdot \boldsymbol{\mathcal{F}}^\rho(\boldsymbol{x}) = \frac{\boldsymbol{\mathcal{J}}^\rho(\boldsymbol{x})^2}{D \rho(\boldsymbol{x})}.
\end{equation}
Interpreting $\boldsymbol{\sigma}^{\rho}$ as a physical dissipation requires that the particle exchanges energy with a thermal reservoir at inverse temperature $\beta$ and satisfies the Einstein relation $\beta D = \mu$.
The second equality follows from a rearrangement of Eq.~\eqref{eq:Fjratio},
\begin{equation}
\boldsymbol{\mathcal{J}}^\rho(\boldsymbol{x}) = D \boldsymbol{\mathcal{F}}^\rho(\boldsymbol{x}) \rho(\boldsymbol{x}),
\end{equation}
which is an expression of linear response; the current at any point in space is linearly proportional to the thermodynamic force at that point.
This linear-response feature of diffusion processes implies that the current fluctuations are specified by local dissipation rates:
\begin{equation}
\mathcal{I}[\rho, \boldsymbol{\mathcal{J}}] = \int d\boldsymbol{x} \ \frac{\boldsymbol{\sigma}^\rho(\boldsymbol{x})}{4 \boldsymbol{\mathcal{J}}^\rho(\boldsymbol{x})^2} \left(\boldsymbol{\mathcal{J}}(\boldsymbol{x}) - \boldsymbol{\mathcal{J}}^\rho(\boldsymbol{x}) \right)^2.
\end{equation}
Our form for $\mathcal{I}$ is analogous to the right-hand side of inequality~\eqref{eq:quadbound}, but for diffusion processes the inequality has become an equality.

\begin{figure}[h!]
\centering
\includegraphics[width=0.45\textwidth]{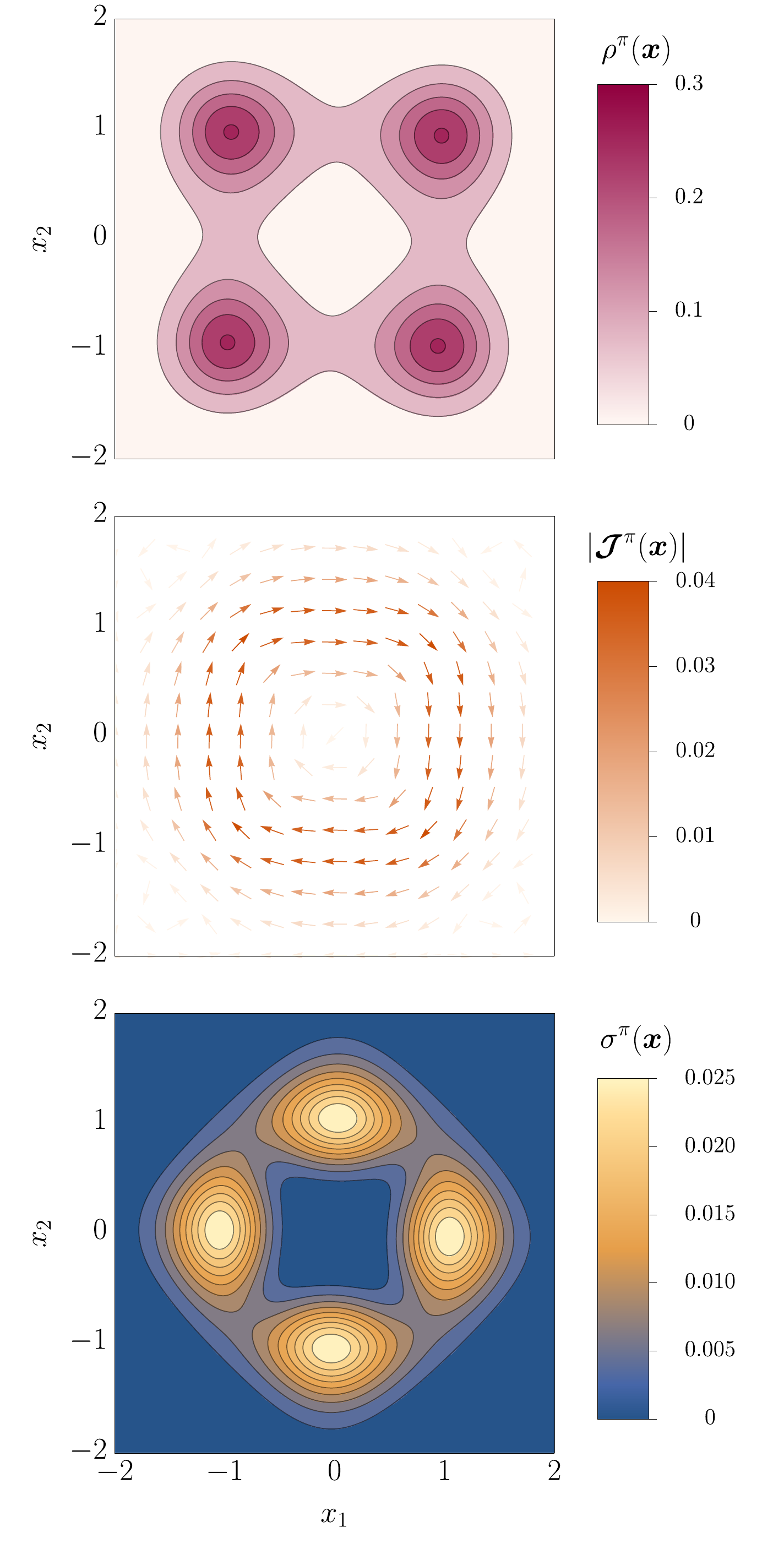}
\caption{Steady-state density field, current field, and local dissipation rate (top to bottom) for the example from Fig.~\ref{fig:diffusionfig} with $A = 8, B = 5, D = 1, \mu = 1$.
The diffusion process was approximated by a jump process on a $400 \times 400$ grid, and the master equation was numerically solved in discrete space.
Note that the local dissipation rate is largest at the barriers, where the density is low and the current high.
}
\label{fig:localdissipationrate}
\end{figure}

\section{Coarse-graining in Space}
\label{sec:macrostate}

Up to this point we have assumed a mesoscopic Markov process that is monitored with complete precision.
More commonly, it is only possible to observe coarse-grained macrostates, and transitions between them might not be Markovian.
Rather, the macroscopic dynamics is described by a hidden Markov model~\cite{Noe2013}.
Remarkably, current fluctuations in the non-Markovian macroscopic dynamics carry information about the dissipation rate of the underlying mesoscopic degrees of freedom.

\begin{figure}
\centering
\includegraphics[width=0.5\textwidth]{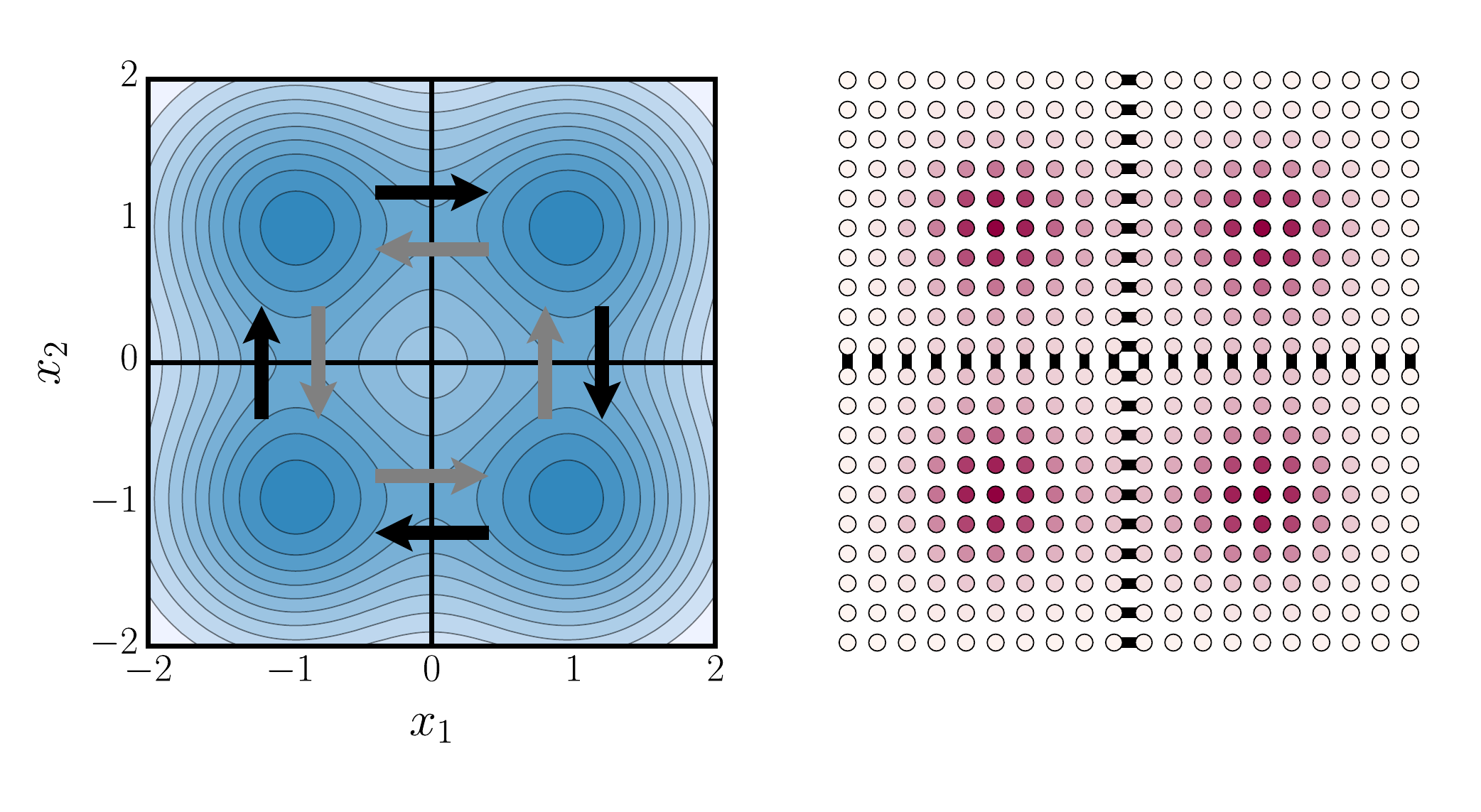}
\caption{Left: The driven diffusion process in Fig.~\ref{fig:diffusionfig} is naturally coarse-grained into four states, and the macroscopic current is constructed by monitoring transitions between them.
Clockwise transitions (black arrows) are given the weight $+1$ while counterclockwise transitions (gray arrows) carry weight $-1$.
All motion within a coarse-grained macrostate is undetected.
Right: In practice, we compute large deviation properties of the coarse-grained current by solving the diffusion process on a grid, as in Sec.~\ref{sec:diffusion}.
On the grid, $j_d$ is constructed by weighting the black edges by $\pm 1$ and all other edges by $0$ as specified in Eq.~\eqref{eq:macrostateddef}.
}
\label{fig:macrostate}
\end{figure}

To illustrate the connection between dissipation and a macroscopic current, we study the generalized scalar current $j_d$, defined in Eq.~\eqref{eq:jddef}.
The vector $\boldsymbol{d}$ determines which microscopic transitions are observable, so for a suitable choice of $\boldsymbol{d}$, $j_d$ can be any macroscopic current.
Consider, for example, the two-dimensional driven diffusion process in Fig.~\ref{fig:diffusionfig}.
A particle evolves on a free energy landscape 
\begin{equation}
U(\boldsymbol{x}) = - B \sum_{i=1}^4 e^{-(\boldsymbol{x} - \boldsymbol{c}_i)^2}
\label{eq:Udef}
\end{equation}
with $B$ controlling the depth of four Gaussian wells centered at $\boldsymbol{c} = (\pm 1, \pm 1)$.
At a coarse-grained level, these wells define four possible macrostates, the quadrants of the coordinate plane.
In addition to the free energy landscape, we introduce a non-gradient external field with amplitude $A$,
\begin{equation}
\boldsymbol{f}_{\rm ext}(\boldsymbol{x}) = A \boldsymbol{x}^2 e^{-3|\boldsymbol{x}|}(x_2, -x_1),
\label{eq:fextdef}
\end{equation}
which drives transitions between the macrostates.
By using Eq.~\eqref{eq:discretizationrates} to discretize the state space, we can numericdally solve for the steady state, as plotted in Fig.~\ref{fig:localdissipationrate}.
We focus on the macroscopic current given by the rate of clockwise (CW) transitions between these coarse-grained states less the rate of counter-clockwise (CCW) transitions.
As illustrated in Fig.~\ref{fig:macrostate}, we measure this current, $j_d = \boldsymbol{j} \cdot \boldsymbol{d}$, by choosing
\begin{equation}
d(y,z) = \begin{cases} 1, & \text{$y\to z$ a CW macrostate transition}\\
-1, & \text{$y\to z$ a CCW macrostate transition}\\
0, & \text{otherwise}.
\end{cases}
\label{eq:macrostateddef}
\end{equation}
The results from Sec.~\ref{sec:scalarcurrent} directly apply to this construction.
In particular, inequality~\eqref{eq:smalldevbound} implies $\Sigma^\pi \geq 2(j_d^\pi)^2 / \text{var}(j_d)$.
The right-hand side of this inequality depends on the first two moments of the \emph{macroscopic} empirical current distribution.
By measuring these moments with coarse-grained observations, we therefore bound the total entropy production of the \emph{mesoscopic} system, $\Sigma^\pi$.

\begin{figure}
\centering
\includegraphics[width=0.45\textwidth]{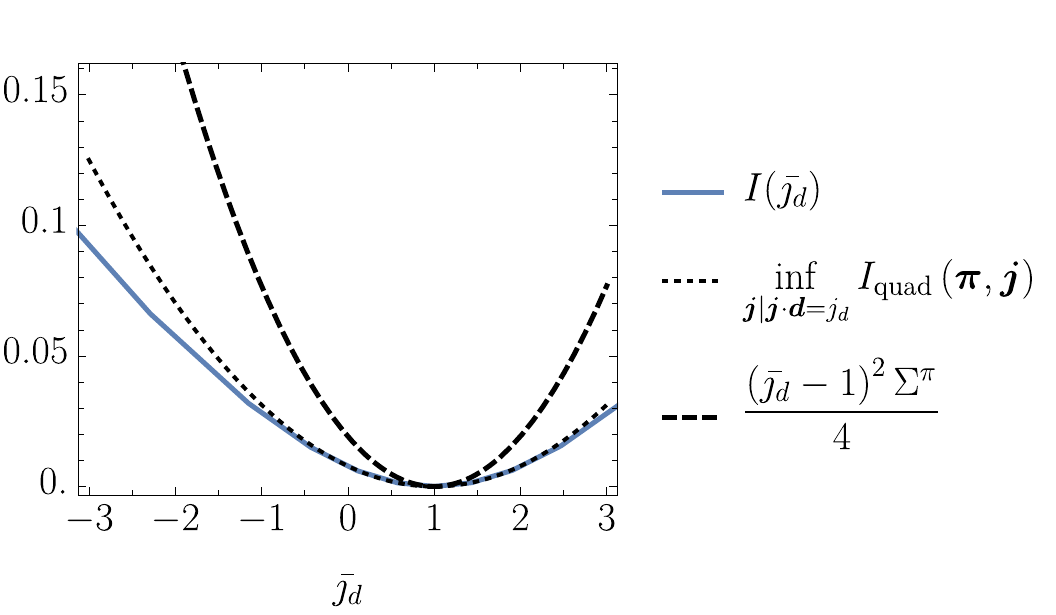}
\caption{
Large deviation function for the macroscopic current with the four coarse-grained states in Fig.~\ref{fig:macrostate}.
$U$ and $\boldsymbol{f}_{\rm ext}$ are given by Eqs.~\eqref{eq:Udef} and~\eqref{eq:fextdef}, respectively, with $A = 8$, $B = 5$.
Macroscopic currents are expressed in relation to the steady-state value: $\bar{\jmath_d} \equiv j_d / j_d^\pi$.
The blue line shows the current fluctuations for the diffusion process solved on a grid of $400 \times 400$ states.
$I(j_d)$ is the numerical Legendre transform of a scaled cumulant generating function, computed as the maximum eigenvalue of a tilted rate matrix~\cite{Lebowitz1999,Touchette2009,Lecomte2007}.
The dotted black line is the result of the constrained minimization of the quadratic form, computed as described in Appendix~\ref{app:tight}.
This constrained minimization very closely approximates $I(\bar{\jmath_d})$ for small deviations.
The dashed black line is the quadratic rate function bound Eq.~\eqref{eq:ratefunctionbound} with curvature regulated by the total dissipation rate.
}
\label{fig:macrostateI}
\end{figure}

\begin{figure}
\centering
\includegraphics[width=0.45\textwidth]{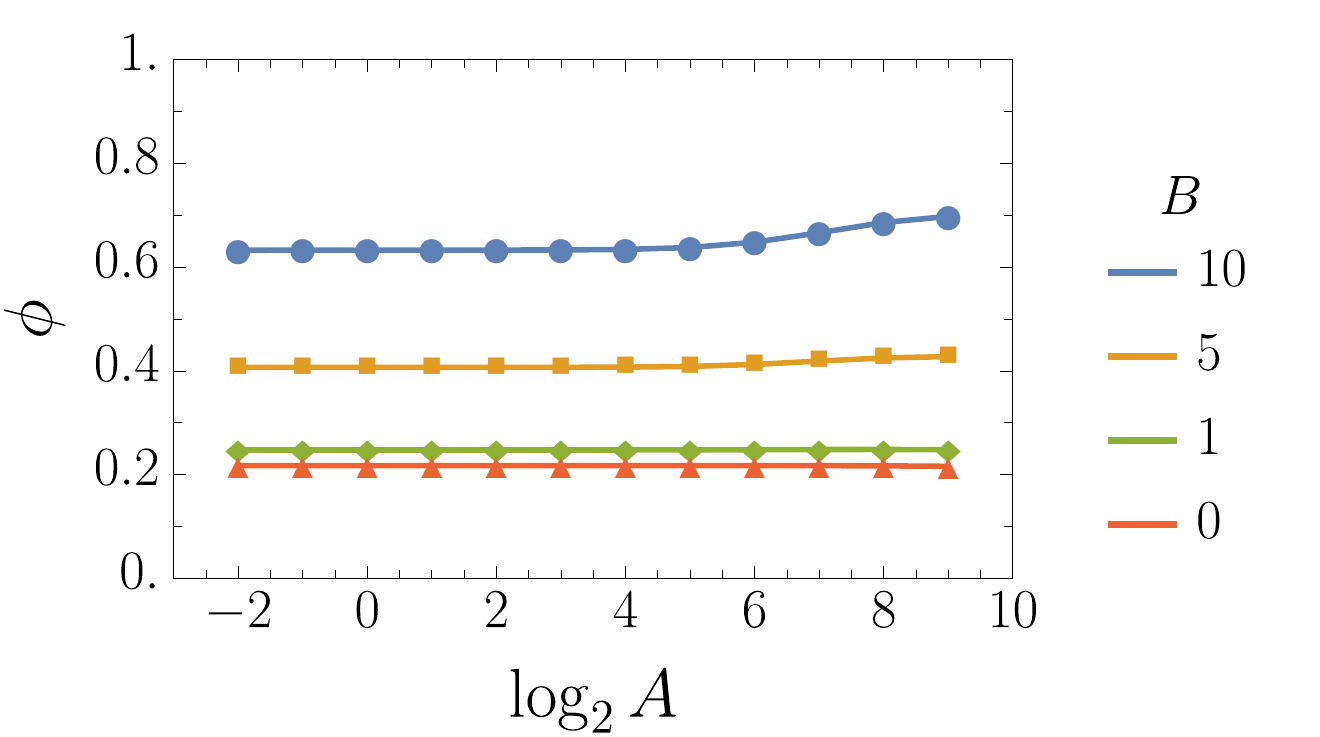}
\caption{
Fraction of the total dissipation rate which can be deduced from the macroscopic current fluctuations in the four-well diffusion model defined by Eqs.~\eqref{eq:Udef} and~\eqref{eq:fextdef}.
$A$ controls the amplitude of the external driving field and $B$ controls the well depths.
Equation~\eqref{eq:dissfrac} defines $\phi$, which takes the value $1$ when the current fluctuations account for all of the dissipation.
Values of $\phi$ were computed by representing the diffusion process as a jump process on a $400 \times 400$ grid.
For large $B$, the wells become deep metastable states which are amenable to coarse-graining, so the macroscopic current accounts for a significant fraction of the dissipation.
}
\label{fig:visiblefraction}
\end{figure}

To assess the inequality's tightness, we used a $400 \times 400$ grid to numerically compute $I(j_d)$ for the four-well model with $D = \mu = 1$, and with various choices of $A$ and $B$.
We saw in Sec.~\ref{sec:diffusion} that $I_{\rm quad}$ becomes exact in the diffusive limit, so
\begin{equation}
I(j_d) = \inf_{\boldsymbol{p}, \boldsymbol{j}|\boldsymbol{j}\cdot \boldsymbol{d} = j_d} I_{\rm quad}(\boldsymbol{p}, \boldsymbol{j}) \leq \inf_{\boldsymbol{j}|\boldsymbol{j}\cdot \boldsymbol{d} = j_d} I_{\rm quad}(\boldsymbol{\pi}, \boldsymbol{j}).
\label{eq:IjdAsinf2}
\end{equation}
The numerical results, plotted in Fig.~\ref{fig:macrostateI}, reflect that inequality~\eqref{eq:IjdAsinf2} is a tight bound for the variance.
In contrast, the dissipation rate bound, inequality~\eqref{eq:ratefunctionbound}, is noticeably weaker.
A measure of this weakness is the ratio
\begin{equation}
\phi = \left(\frac{2(j_d^\pi)^2}{\text{var}(j_d)}\right) / \Sigma^\pi,
\label{eq:dissfrac}
\end{equation}
which ranges from $0$ to $1$.
This ratio, plotted in Fig.~\ref{fig:visiblefraction} for a range of driving amplitudes and well depths, assesses the fraction of the dissipation rate that can be inferred from the macroscopic current fluctuations.

It has been shown that our inference strategy detects the greatest fraction of dissipation if the macroscopic current is constructed to be proportional to the thermodynamic force~\cite{Gingrich2016}.
More precisely, $\phi = 1$ when $\boldsymbol{d} \propto \boldsymbol{F}^\pi$, and $\phi$ decreases when the vector $\boldsymbol{d}$ is less aligned with $\boldsymbol{F}^\pi$.
In our coarse-graining example, the elements of $\boldsymbol{d}$ are set to $0$ or $\pm 1$ based on whether each mesoscopic transition is observable.
With $\boldsymbol{d}$ constrained by which transitions are observable, $\phi$ can only be varied if $\boldsymbol{F}^\pi$ is altered, something that can be achieved by tuning $A$ and $B$.

\begin{figure}
\centering
\includegraphics[width=0.49\textwidth]{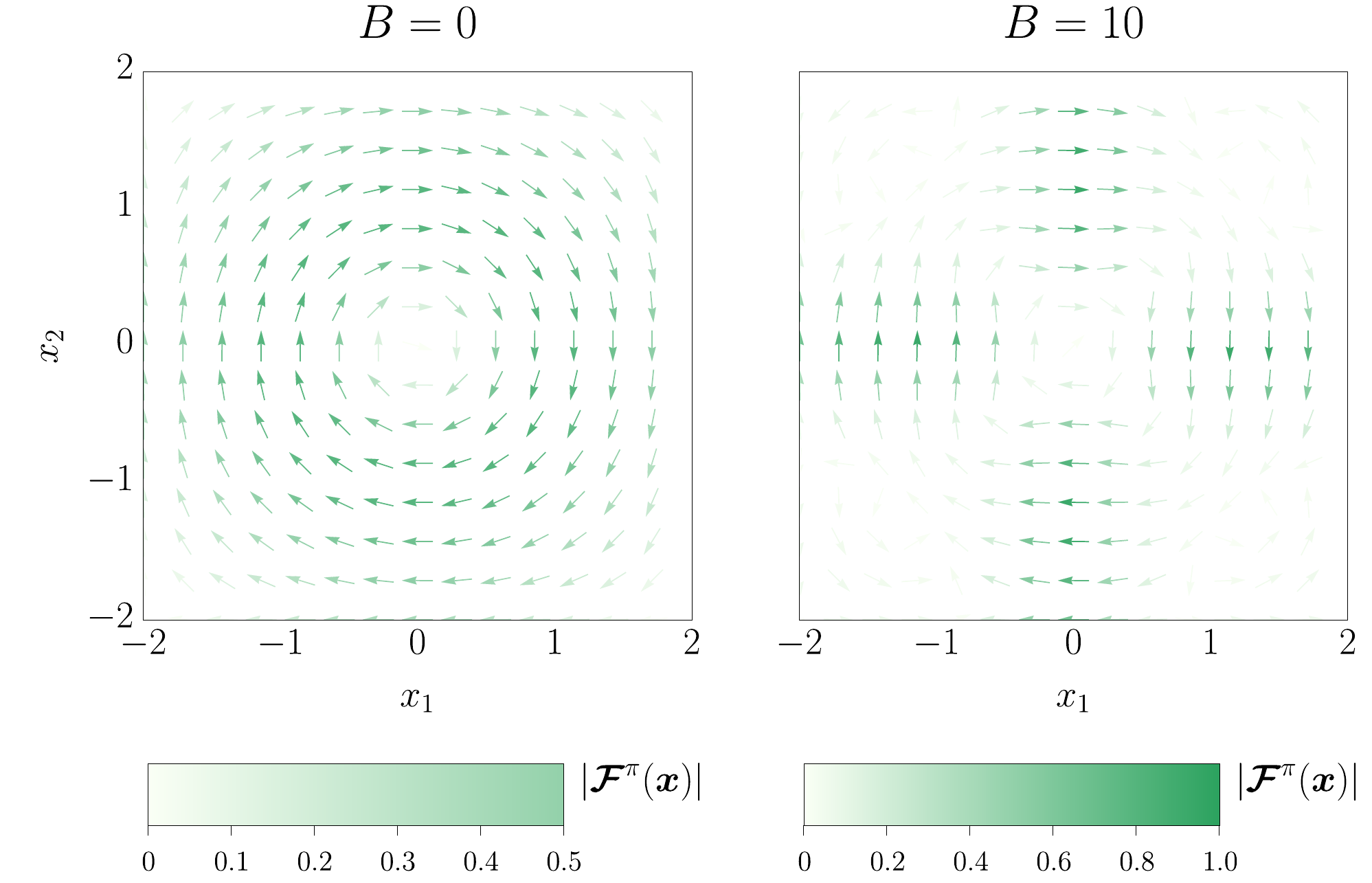}
\caption{
Thermodynamic force for the four-well model with $A = 8$, solved on a $400 \times 400$ grid.
For large $B$, the free energy landscape has deep wells, and the thermodynamic forces are large along the barriers separating the wells.
In this large-$B$ limit, the thermodynamic force resembles $\boldsymbol{d}$, which vanishes everywhere except for along the axes.
}
\label{fig:thermodynamicforces}
\end{figure}

We can understand $\phi$'s dependence on these model parameters by analyzing the degree to which $\boldsymbol{F}^\pi$ aligns with $\boldsymbol{d}$.
Fig.~\ref{fig:thermodynamicforces} illustrates how the continuous thermodynamic force field $\boldsymbol{\mathcal{F}}^\pi(\boldsymbol{x})$ varies with the well depth.
In the absence of wells ($B = 0$), the thermodynamic force is radially symmetric about the origin.
As the wells are made deeper, $\boldsymbol{\mathcal{F}}^\pi$ is amplified along the axes, more closely resembling $\boldsymbol{d}$ of Fig.~\ref{fig:macrostate}.
This resemblance explains the trend that $\phi$ increases with increasing $B$.
Interestingly, $\phi$ depends only weakly on the external field amplitude, a fact we rationalize with a linear-response argument at the end of Appendix~\ref{app:tight}.

\section{Conclusions}
\label{sec:conclusions}
It is typically impossible to resolve detailed mesoscopic dynamics, even with the most sophisticated experimental tools.
Because the dissipation rate is defined in terms of such immeasurable mesoscopic transitions, it may seem that there is no hope of experimentally measuring the dissipation, short of explicitly detecting the free energetic flows from the thermodynamic reservoirs.
Our models and analysis demonstrate that, in fact, a bound on the total dissipation can be inferred by monitoring only the fluctuations in macroscopic currents.
The procedure we use to infer this bound is extremely adaptable; we make essentially no assumptions about the mesoscopic details of the system nor about the macroscopic currents that will be observed.

Crucially, it is the macroscopic current fluctuations, not their averages, that reveal the dissipation of the unobserved mesoscopic dynamics.
Motivated by the inability to fully resolve mesoscopic dynamics in an experiment, we have constructed a coarse-grained macroscopic current by recording a small subset of the mesoscopic transitions.
The fraction of total dissipation due to the irreversibility of one of these observed transitions is small and in fact vanishes in the diffusive limit.
Were we to measure only the average current through the observed links, we would at best deduce this infinitesimally small portion of the total dissipation.
However, if we use the fluctuation-dissipation relation for the observed currents, we tacitly constrain the average behavior of all the unobserved transitions.
Hence the fluctuations in a tiny subset of the mesoscopic transitions carry information about an appreciable fraction of the total dissipation.
This useful fact stems from a fundamental bound on the extent of fluctuations: the total dissipation rate sets a bound for the variance in \emph{any} generalized current.

We anticipate that our inference scheme will be a robust strategy for analyzing dissipation in stochastic, biophysical systems.
The extent to which coarse-grained fluctuations reveal the total dissipation, measured by $\phi$, depends on the fidelity of the coarse-graining.
For the model we studied in Sec.\ref{sec:macrostate}, $\phi$ nears $1$ when the wells are sufficiently deep that the coarse-grained states are long-lived metastable macrostates.
The step-like dynamics of dissipative biological machines~\cite{Smith2001,Hayashi2010} demonstrate precisely this type of metastability and time-scale separation~\cite{Wang2016}. 
Indeed, it is often the goal of biological studies to assign the experimentally observed metastable macrostates to coarse-grained descriptions, e.g., ligation states.
Together, these observations bolster the prospect of accurately inferring dissipation rates using data from single molecule experiments.

\begin{acknowledgments}
We wish to thank an anonymous reviewer for useful feedback pertaining to Section~\ref{sec:heuristic}.
This research is funded by the Gordon and Betty Moore Foundation to TRG as a Physics of Living Systems Fellow through Grant GBMF4513 and to JMH through Grant GBMF4343.
GMR was supported by a National Science Foundation Graduate Research Fellowship.
\end{acknowledgments}

\appendix
\section{Derivation of equivalent forms of $\Psi$}
\label{app:ratefnderivation}
In the main text we have extensively discussed the time-asymmetric empirical current from state $y$ to $z$, $j(y,z)$.
This quantity is time-asymmetric because it acquires a negative sign if the trajectory is run backwards in time.
Maes and coworkers have highlighted the importance of the time-symmetric empirical traffic, which counts the total number of hops across the $yz$ edge in either direction~\cite{Maes2007,Maes2008,Maes2008On,Baiesi2009,Maes2016}.
As in our study of the currents, we monitor two different forms of the traffic: the expected traffic given $\boldsymbol{p}$, $t^p(y,z) = p(y)r(y,z) + p(z)r(z,y)$, and the traffic $t_\star(y,z) = q_\star(y,z) + q_\star(z,y)$.
$\Psi$ naturally decomposes into contributions from the traffic and contributions from the currents, $\Psi = \Psi_{\rm traffic} + \Psi_{\rm current}$~\cite{Maes2008},
with
\begin{equation}
\nonumber \Psi_{\rm traffic} \equiv p(y)r(y,z) + p(z) r(z,y) - q_\star(y,z) - q_\star(z,y)
\end{equation}
and
\begin{equation}
\Psi_{\rm current} \equiv q_\star(y,z) \ln \frac{q_\star(y,z)}{p(y) r(y,z)} + q_\star(z,y) \ln \frac{q_\star(z,y)}{p(z) r(z,y)}.
\end{equation}

$\Psi_{\rm traffic}$ is simply the difference between the expected and the observed number of hops, $\Psi_{\rm traffic} = t^p(y,z) - t_\star(y,z)$.
To bring $\Psi_{\rm traffic}$ into the form that appears in Eq.~\eqref{eq:psibertini}, we recognize that $t_\star(y,z) = \sqrt{j(y,z)^2 + a^p(y,z)^2}$ and that $t^p(y,z) = \sqrt{j^p(y,z)^2 + a^p(y,z)^2}$.
Simplifying $\Psi_{\rm current}$ is slightly more involved.
Using the solution for $q_\star$, Eq.~\eqref{eq:qstar}, $\Psi_{\rm current}$ can be rewritten in the form of Eq.~\eqref{eq:psibertini},
\begin{align}
\nonumber \Psi_{\rm current} &= \frac{j(y,z)}{2} \ln \frac{q_\star(y,z) p(z) r(z,y)}{q_\star(z,y) p(y) r(y,z)}\\
\nonumber &\ \ \ + \frac{\sqrt{j(y,z)^2 + a^p(y,z)^2}}{2} \underbrace{\ln \frac{q_\star(y,z) q_\star(z,y)}{p(y) r(y,z) p(z) r(z,y)}}_{0}\\
\nonumber &= \frac{j(y,z)}{2} \left(\ln \frac{q_\star(y,z)}{q_\star(z,y)} + \ln \frac{p(z) r(z,y)}{p(y) r(y,z)}\right)\\
\nonumber &= \frac{j(y,z)}{2} \ln \frac{j(y,z) + \sqrt{j(y,z)^2 + a^p(y,z)^2}}{-j(y,z) + \sqrt{j(y,z)^2 + a^p(y,z)^2}}\\
\nonumber &\ \ \ + \frac{j(y,z)}{2} \ln \frac{j^p(y,z) + \sqrt{j^p(y,z)^2 + a^p(y,z)^2}}{-j^p(y,z) + \sqrt{j^p(y,z)^2 + a^p(y,z)^2}}\\
&= j(y,z)\left(\arcsinh\frac{j(y,z)}{a^p(y,z)} - \arcsinh\frac{j^p(y,z)}{a^p(y,z)}\right).
\end{align}

\section{Tight Quadratic Current Fluctuation Bound}
\label{app:tight}
In the main text, an upper bound for $I(j_d)$ was obtained by evaluating $I$ at $\boldsymbol{p}^* = \boldsymbol{\pi}$ and $\boldsymbol{j}^* = (j_d / j_d^\pi) \boldsymbol{j}^{\pi}$.
This choice implies that the chance of measuring a value $j_d$ is at least as likely as it would be if we simultaneously scaled all of the steady-state mesoscopic currents so as to make $\boldsymbol{j}^* \cdot \boldsymbol{d} = j_d$.
There can, however, be more likely ways to obtain $j_d$ which do not equally scale the different mesoscopic currents.
Identifying the optimum is a straightforward exercise in linear algebra: the quadratic form $I_{\rm quad}$ must be extremized subject to the linear constraint $\boldsymbol{j} \cdot \boldsymbol{d} = j_d$ and to current conservation.
The restriction to conservative currents may be imposed by a collection of Lagrange multipliers, one per node, as in Ref.~\cite{Gingrich2016} or by expressing the currents in a cycle basis as in Ref.~\cite{Polettini2016,Pietzonka2016}.
We use the second strategy here.
Since both strategies exactly solve the same convex optimization problem, they must yield the same solution.

Let $a(i)$ be the current around the $i^{\rm th}$ cycle in a complete cycle basis.
This basis may be constructed from spanning trees~\cite{Schnakenberg1976}, but for now the particular cycle basis is immaterial. 
We introduce
\begin{equation}
\chi^i_{yz} = \begin{cases} 1, & \text{cycle }i\text{ contains edge }yz\text{ with }y<z\\
-1, & \text{cycle }i\text{ contains edge }yz\text{ with }z<y\\
0, & \text{otherwise}
\end{cases}
\end{equation}
to convert between the edge basis and the cycle basis so that
\begin{equation}
j(y,z) = \sum_i \chi^i_{yz} a(i).
\end{equation}
Recall that
\begin{equation}
I_{\rm quad}(\boldsymbol{p}, \boldsymbol{j}) = \begin{cases} 
\sum\limits_{y<z} \frac{\sigma^p(y,z)}{4} \left(\frac{j(y,z)}{j^p(y,z)} - 1\right)^2,& \text{conservative }\boldsymbol{j}\\
\infty, & \text{otherwise}.
\end{cases}
\end{equation}
In terms of the cycle basis the restriction to conservative currents is more natural:
\begin{align}
\nonumber I_{\rm quad}(\boldsymbol{p}, \boldsymbol{a}) &= \sum_{y<z} \left(\sum_i \chi^i_{yz} (a(i) - a^p(i))\right)^2\frac{F^p(y,z)}{4 j^p(y,z)}\\
&=\frac{1}{2} \sum_{i,k} (a(i) - a^p(i)) G_{ik} (a(k) - a^p(k)),
\end{align}
with $G_{ik} = \sum_{y<z} \chi^i_{yz} \chi^k_{yz} \frac{F^p(y,z)}{2 j^p(y,z)}$~\cite{Polettini2016}.
We also translate the linear constraint into the cycle basis:
\begin{equation}
j_d = \sum_{y<z} \sum_i \chi^i_{yz} a(i) d(y,z) = \sum_i a(i) d(i) \equiv \boldsymbol{a} \cdot \boldsymbol{d},
\end{equation}
where we have defined $d(i) \equiv \sum_{y<z} \chi^i_{yz} d(y,z)$.
In words, $d(i)$ is now the expansion coefficient that indicates how much cycle $i$ contributes to the generalized current $j_d$.

In analogy with the main text, $I(j_d) \leq I_{\rm quad}(\boldsymbol{p}^*, \boldsymbol{a}^*)$ for any choice of $\boldsymbol{p}^*$ and $\boldsymbol{a}^*$ which satisfy the constraint $\boldsymbol{a} \cdot \boldsymbol{d} = j_d$.
We again choose $\boldsymbol{p}^* = \boldsymbol{\pi}$, but now we compute the exact minimizer of the quadratic form $I_{\rm quad}$:
\begin{equation}
\boldsymbol{a}^* = \boldsymbol{a}^\pi + \left(\frac{j_d - j_d^\pi}{\boldsymbol{d}^T G^{-1} \boldsymbol{d}}\right)G^{-1} \boldsymbol{d}.
\end{equation}
Hence,
\begin{equation}
I(j_d) \leq \inf_{\boldsymbol{j}|\boldsymbol{j}\cdot \boldsymbol{d} = j_d} I_{\rm quad}(\boldsymbol{\pi}, \boldsymbol{j}) = \frac{\left(j_d - j_d^\pi\right)^2}{2 \boldsymbol{d}^T G^{-1} \boldsymbol{d}}.
\end{equation}

For the coarse-graining problem of Sec.~\ref{sec:macrostate}, this denominator is simple if we use the cycle basis consisting of the square lattice's plaquettes.
Each plaquettes is one of the squares of side length $h$.
Notice that for our coarse-graining procedure, $d(i) = 0$ for all but one of the plaquettes.
Only the central plaquette that encircles the origin has a nonzero $d$.
Since we can order the plaquettes arbitrarily, we designate the central plaquette by the label ``o'' to highlight that it is the plaquette at the origin.
Because $d({\rm o}) = 4$, $\boldsymbol{d}^T G^{-1} \boldsymbol{d} = 16 G^{-1}_{\rm oo}$ with $G^{-1}_{\rm oo}$ denoting the diagonal matrix element of $G^{-1}$ for the plaquette at the origin.

Finally, we can see why $\phi$ is nearly independent of the driving field amplitude $A$ in Fig.~\ref{fig:visiblefraction}.
In terms of the $G$ matrix,
\begin{equation}
\phi = \left(\frac{2(j_d^\pi)^2}{\text{var}(j_d)}\right) / \Sigma^\pi = \frac{(j_d^\pi)^2}{8 G^{-1}_\text{oo} \Sigma^\pi}.
\end{equation}
For sufficiently small driving amplitude, we anticipate that the mesoscopic currents will respond linearly to the external force: $\boldsymbol{j}^\pi \propto A$.
In this linear regime we also have $\boldsymbol{F}^\pi \propto A$, so $\Sigma^\pi \propto A^2$.
Furthermore, since matrix elements of $G$ involve ratios of $F^p$ and $j$, the linear-response value of $G^{-1}_\text{oo}$ is independent of $A$.
Combining these effects, we see that $\phi$ does not vary with $A$ in the linear-response regime.
Empirically, we observe in Fig.~\ref{fig:visiblefraction} that insensitivity to $A$ in fact extends far beyond the linear-response regime.
\bibliography{refs}
\end{document}